\documentclass[twocolumn]{aastex61}

\usepackage{amsmath}
\usepackage{amsfonts}
\usepackage{amssymb}

\received{2017 April 12}
\revised{2017 May 24}
\accepted{2017 May 25}
\submitjournal{ApJ}
\shorttitle{Search for VHE Emission from GRBs with HAWC}
\shortauthors{HAWC Collaboration et al.}

\begin{document}
\title{Search for Very-High-Energy Emission from Gamma-Ray Bursts Using the First 18 Months of Data from the HAWC Gamma-Ray Observatory}

\correspondingauthor{Dirk Lennarz}
\email{dirk.lennarz@gatech.edu}

\author{R.~Alfaro}
\affiliation{Instituto de F\'{i}sica, Universidad Nacional Aut\'{o}noma de M\'{e}xico, Ciudad de M\'{e}xico, Mexico}

\author{C.~Alvarez}
\affiliation{Universidad Aut\'{o}noma de Chiapas, Tuxtla Guti\'{e}rrez, Chiapas, Mexico}

\author{J.D.~\'{A}lvarez}
\affiliation{Universidad Michoacana de San Nicol\'{a}s de Hidalgo, Morelia, Mexico}

\author{R.~Arceo}
\affiliation{Universidad Aut\'{o}noma de Chiapas, Tuxtla Guti\'{e}rrez, Chiapas, Mexico}

\author{J.C.~Arteaga-Vel\'{a}zquez}
\affiliation{Universidad Michoacana de San Nicol\'{a}s de Hidalgo, Morelia, Mexico}

\author{D.~Avila Rojas}
\affiliation{Instituto de F\'{i}sica, Universidad Nacional Aut\'{o}noma de M\'{e}xico, Ciudad de M\'{e}xico, Mexico}

\author{H.A.~Ayala Solares}
\affiliation{Department of Physics, Michigan Technological University, Houghton, MI, USA}

\author{A.S.~Barber}
\affiliation{Department of Physics and Astronomy, University of Utah, Salt Lake City, UT, USA}

\author{N.~Bautista-Elivar}
\affiliation{Universidad Politecnica de Pachuca, Pachuca, Hidalgo, Mexico}

\author{A.~Becerril}
\affiliation{Instituto de F\'{i}sica, Universidad Nacional Aut\'{o}noma de M\'{e}xico, Ciudad de M\'{e}xico, Mexico}

% ...

\collaboration{(The HAWC collaboration)}
\noaffiliation

\begin{abstract}
The High Altitude Water Cherenkov (HAWC) Gamma-ray Observatory is an extensive air shower detector operating in central Mexico, which has recently completed its first two years of full operations. If for a burst like GRB~130427A at a redshift of 0.34 and a high-energy component following a power law with index −1.66, the high-energy component is extended to higher energies with no cut-off other than from extragalactic background light attenuation, HAWC would observe gamma rays with a peak energy of $\sim$300~GeV. This paper reports the results of HAWC observations of 64 gamma-ray bursts (GRBs) detected by \emph{Swift} and \emph{Fermi}, including three GRBs that were also detected by the Large Area Telescope (\emph{Fermi}-LAT). An ON/OFF analysis method is employed, searching on the time scale given by the observed light curve at keV--MeV energies and also on extended time scales. For all GRBs and time scales, no statistically significant excess of counts is found and upper limits on the number of gamma rays and the gamma-ray flux are calculated. GRB~170206A, the third brightest short GRB detected by the Gamma-ray Burst Monitor on board the \emph{Fermi} satellite (\emph{Fermi}-GBM) and also detected by the LAT, occurred very close to zenith. The LAT measurements can neither exclude the presence of a synchrotron self-Compton (SSC) component nor constrain its spectrum. Instead, the HAWC upper limits constrain the expected cut-off in an additional high-energy component to be less than $100~\rm{GeV}$ for reasonable assumptions about the energetics and redshift of the burst.
\end{abstract}


\section{Introduction}
After the discovery of gamma-ray bursts \citep[GRBs;][]{bib:GRB_discovery} more than a hundred models have been proposed to explain their origins \citep[e.g.][]{bib:GRB_model_number}. Despite these efforts and a lot of progress in recent years, the details of the emission mechanisms are still poorly understood. GRBs have a prompt emission phase that typically lasts from $\sim10^{-2}\nobreakspace\rm{s}$ to $\sim10^{3}\nobreakspace\rm{s}$. The light curve in the keV--MeV range is irregular, diverse and complex, with no two GRBs being alike. After the prompt phase, the X-ray light curve becomes more regular and follows a ''canonical'' light curve that is comparable between bursts \citep{bib:GRB_early_afterglow}. It transitions into a long and smooth decay, which can be studied using multi-wavelength observations. The emission of GRBs can be explained by a collimated, relativistic outflow \citep[the so-called fireball model; for a review see e.g.][]{bib:Piran_fireball_review}. Relativistic ejecta propagating at various velocities can collide in the outflow and create internal shocks. The interaction of outflowing material with the surrounding medium can create external shocks (forward and reverse).

GRBs are classified into two main groups according to the duration of their prompt emission, which is measured as $T_{90}$, the central time interval in which 90\% of the total fluence is detected \citep{bib:GRB_long_short}. The duration of long GRBs is greater than two seconds, while short GRBs are shorter than two seconds. Since some long GRBs are also associated with Type Ib/Ic core-collapse supernovae \citep{bib:GRB_SNe}, the core collapse of a rapidly rotating star \citep[collapsar;][]{bib:collapsar} is a likely candidate for long GRBs. Short GRBs are generally attributed to the merger of two compact stellar remnants \citep[double neutron stars or neutron star--black hole binaries;][]{bib:short_GRB_origin_1,bib:short_GRB_origin_2}. Both source candidates can create the central engine required to start the relativistic outflow.

The GRB prompt emission features a non-thermal spectrum that is, for the most part, well described by the phenomenological Band function, which is composed of two smoothly connected power laws peaking at $\lesssim$~MeV energies \citep{bib:band_function}. However, in recent years, evidence has grown that two additional components may be present besides the Band function \citep[e.g.][and references therein]{bib:GRB_spectral_components}: a thermal component, peaking typically at $\sim$100~keV, and a component at high energies ($\gtrsim$100~MeV) with a completely separate temporal evolution. The high-energy component has a delayed onset with respect to emission at lower energies and features a long-lived emission ($\gtrsim$ 100~s), which decays smoothly as $\sim t^{-1}$ as a function of time $t$ \citep{bib:Fermi_LAT_GRB_catalogue}.

The Band function spectra are often attributed to the synchrotron emission of internal-shock accelerated electrons. The maximum photon energy of synchrotron emission is limited \citep[e.g.][]{bib:max_synchrotron}, but in so-called leptonic models, the internal shock could produce high-energy emission through synchrotron self-Compton (SSC) processes \citep[e.g.][]{bib:internal_shock_SSC}. Other internal-shock models, the so-called hadronic models, consider proton synchrotron emission \citep[e.g.][]{bib:internal_shock_proton_synchrotron} and/or the inverse Compton emission from secondary particles produced in internal cascades \citep{bib:internal_shock_proton_cascades}.

The GRB afterglow emission is attributed to the external shocks, and in the external-shock scenario they could also be the origin of the high-energy component. In the very bright GRB~130427A, several photons were found that appear to be incompatible with the model of synchrotron radiation emitted by electrons accelerated at the external forward shock \citep{bib:GRB130427A_Fermi_Science}. In leptonic scenarios, possible origins of the high-energy component are SSC emission in the forward shock \citep[e.g.][]{bib:external_fs_shock_SSC} or the reverse shock \citep[e.g.][]{bib:external_rs_shock_SSC} or a combination of both \citep[e.g.][]{bib:external_shocks_SSC}. Hadronic models also exist for the external-shock scenario \citep[see e.g.][]{bib:external_shocks_hadrons}.

The observation and temporal evolution of GRB spectra at very-high energies (VHE; $>100\nobreakspace\rm{GeV}$) can help distinguish between the myriad of models. For example, the detection of TeV photons at early times would be difficult to explain with internal SSC emission. Satellite experiments, like the Large Area Telescope on board the \emph{Fermi} spacecraft \citep[\emph{Fermi}-LAT;][]{bib:LAT_instrument}, are limited in size and, since the photon flux decreases steeply with energy, only a few photons can be expected in the VHE regime. Imaging Atmospheric Cherenkov Telescopes (IACTs) are sensitive in this energy range and routinely look for TeV emission from GRBs, but only upper limits have been reported so far \citep{bib:MAGIC_GRB,bib:HESS_GRB,bib:VERITAS_GRB}. One limitation for IACTs is that they are pointed instruments and thus need to slew the telescopes to the GRB position first. They are therefore not likely to observe long GRBs during their prompt phase, but will mostly observe in the afterglow phase \citep[an exception to this can be found in][]{bib:HESS_GRB_PROMPT}. Observational constraints lead to a $\sim$10\% duty cycle, and the Cherenkov Telescope Array (CTA), the next-generation IACT array, is expected to detect of the order of a few GRBs per year \citep{bib:CTA_GRB_estimate}.

The other ground-based technique uses extensive air shower (EAS) arrays that feature a large field of view ($\sim2\nobreakspace\rm{sr}$ or 16\% of the sky) and near 100\% duty cycle. They are less sensitive to transient sources than IACTs, but the lack of observational delays allows them to observe the prompt phase of long and short GRBs. The High Altitude Water Cherenkov (HAWC) Observatory is a new EAS array (see next section). HAWC is mostly sensitive to short GRBs and can detect as many as 1 to 2 GRBs per year \citep{bib:HAWC_GRB_rate}. However, this study used a sensitivity at low energies that has not been achieved yet. In this paper, the results of the analysis of GRB data with the current best analysis from the first 18 months of full operation of HAWC are reported.

\section{The HAWC Gamma-Ray Observatory}
HAWC is located at Sierra Negra, Mexico, at an altitude of 4,100~m above sea level \citep{bib:HAWC_crab}. It employs the water Cherenkov technique, in which VHE photons are detected by measuring Cherenkov light from secondary particles created in the EAS. HAWC comprises of 300 steel tanks containing light-proof bladders of 7.3 m diameter and 4.5 m depth, each holding $\sim$200,000 litres of filtered water. Each tank is instrumented with three $8^{\prime\prime}$ photomultiplier tubes (PMTs) and one $10^{\prime\prime}$ PMT on the bottom. Cables connect the PMTs to a central counting house, where the PMT pulses are shaped and discriminated at two thresholds.

HAWC features two systems for data acquisition (DAQ). The main DAQ measures the arrival time and time over threshold (TOT) of PMT pulses. The TOT is characteristic of the PMT's measurement of the charge (and hence the amount of Cherenkov light in the tank) and the information from different tanks allows the reconstruction of the EAS core. The core and the signal arrival time in different tanks permit the reconstruction of the direction of the incident shower. The scaler DAQ counts each PMT signal that exceeds the lower discrimination threshold in $100\nobreakspace\rm{ms}$ windows. GRBs are detected by a statistically significant excess of counts over the noise rate \citep[``single particle technique'';][]{bib:scaler_method}. The scaler DAQ is sensitive to showers of lower energies that cannot be reliably reconstructed with the main DAQ. These different energy sensitivities make the DAQs complement each other \citep{bib:HAWC_GRBs}. In this paper, data from the main DAQ are analysed.

HAWC uses a software trigger. All PMT signals are kept in memory and an EAS event is formed if a certain threshold on the number of coincident signals in a 150~ns time window is met (trigger threshold). The overwhelming majority of events recorded by HAWC originate from hadronic cosmic rays. The lateral charge distribution on the ground can help to determine the species of the primary particle and a set of cuts, called the gamma--hadron (GH) separation cuts (see further information in the next section), can be used to obtain a purer gamma-ray event sample.

Construction of the HAWC Observatory began in 2012. During the early construction phase, when 29 tanks with 115 deployed PMTs were operational (HAWC-30), HAWC provided the first limits on the prompt VHE emission of GRB~130427A, the most powerful burst ever detected with a redshift $z\lesssim0.5$  \citep{bib:HAWC_GRB130427A}. The small detector and large zenith angle of the GRB caused the limits to be about two orders of magnitude higher than a simple extrapolation of the \emph{Fermi} data.

Science operations of the HAWC array started with a partially built array (called HAWC-111) in August 2013. A search for VHE emission from 18 \emph{Swift}-detected GRBs that occurred in the HAWC field of view between 2013 August 2 and 2014 July 8 showed that none of the GRBs is significant above $3~\sigma$ after accounting for trial factors \citep{bib:HAWC-111_GRBs}.

Data taking with 250 tanks began on 2014 November 26 and even though the array was still gradually growing, this data set is regarded as full HAWC data. In this paper, GRBs observed up to 2016 June 29 are covered. During this period, detector uptime was about 92\%, with downtime mainly caused by power failures and scheduled construction and maintenance \citep{bib:HAWC_crab}. During the time from HAWC-111 to the beginning of the full operation of HAWC, three bursts detected by \emph{Fermi}-LAT, one burst detected by \emph{Swift} and one burst triangulated by the Interplanetary Network (IPN) were in the HAWC field of view, but none of the GRBs was significant above $2~\sigma$ \citep{bib:HAWC_GRBs_Gamma}.

HAWC has an extensive GRB programme. An automated analysis is triggered at the HAWC site, if a GRB reported by satellites through the GRB Coordinates Network\footnote{\url{http://gcn.gscfc.nasa.gov}} (GCN) is within the HAWC field of view (for zenith angles smaller than $45\arcdeg$). Results are available within $\sim$30~minutes. The analysis can also be redone if more accurate information becomes available later. HAWC reported the result of the analysis of GRB~160509A, a very bright burst detected by \emph{Fermi}-LAT, but at a very high zenith angle of $62\arcdeg$ at the HAWC site, within two days after the burst \citep{bib:HAWC_GCN_GRB160509A}. HAWC also runs an untriggered search of the entire overhead sky \citep[for more information, see][]{bib:HAWC_GRBs_Gamma}.

In addition to the analysis at the HAWC site, GRBs are also analysed off-site. This allows the use of improved calibration, event reconstruction and analysis methods, and all the available information on a certain GRB. In this paper, results from the off-site GRB analysis are shown.

\section{Analysis Method}
HAWC classifies events into 10 size bins (zero to nine) based on the fraction of the PMTs that participate in the event. This method yields more stable results than cutting on the absolute number of PMTs in an event, as it compensates for detector changes, such as in the final stages of construction. The current trigger threshold of HAWC allows for events below the lowest size bin (here called threshold events). Events in higher size bins correspond on average to higher energy primary particles and have a better point spread function (PSF) and a better GH separation.

The analysis presented here is based on HAWC observations of the Crab Nebula \citep{bib:HAWC_crab}. GH separation cuts were found by optimising the statistical significance of the Crab using about 60\% of the data set. In size bins 1 to 9, the cut variables are, within systematic uncertainties, well modelled using Monte Carlo simulations of gamma- and cosmic-ray air showers and a detailed simulation of the detector. There is a discrepancy in size bin 0 and, at the current time, no reliable set of GH separation cuts is available for this bin. This is why the events in bin 0 and the threshold events were excluded from the Crab analysis. Details on the sensitivity, cuts, background rejection and systematic uncertainties can be found in \cite{bib:HAWC_crab}.

The Crab analysis uses a binned-likelihood method, but here a classical ON/OFF method \citep[for example used in the Whipple observatory;][]{bib:ON_OFF} is employed. In short, it consists of defining a search circle around the position of the GRB, determining an estimated number of background events that will appear in it and finally calculating the number of events above background in the search circle. The $p$-value, which is the probability of obtaining equal or more counts from the background than the ones in the search circle, is calculated assuming a Poisson distribution with the estimated background as the mean.

\subsection{Event Selection}
In this paper, only a single analysis bin is used, combining size bins 1 to 9 with the GH separation cuts corresponding to size bin 1 of the Crab analysis. VHE photons are attenuated due to interactions with the extragalactic background light (EBL), so most of the signal from a GRB is expected at low energies. Hence, the expected signal is dominated by the lowest bin, so the gain from a multi-bin analysis with separate GH separation cuts for each bin is expected to be small. In previous analyses, the events were selected by combining size bins 0 to 9 plus the threshold events and applying no GH separation cuts \citep{bib:HAWC_GRBs_Gamma}.

The efficiency of the event selections can be compared for exceptional, historic GRBs at different redshifts. EBL absorption is assumed according to the fiducial model in \cite{bib:EBL}. For a nearby burst like GRB~130427A ($z=0.34$), where the high-energy component followed a power law with index $-1.66$ \citep{bib:GRB130427A_Fermi_Science}, the analysis starting at bin 1 with GH separation cuts nearly doubles the expected significance compared to an analysis using bins 0--9 and no GH separation cuts. For a short burst like GRB~090510 at a redshift of 0.903 and a high-energy component proportional to $E^{-1.54}$ \citep{bib:090510_Fermi}, the two event selections result in a similar significance. At higher redshifts, e.g. GRB~090902B at a redshift of 1.8 and a high-energy component proportional to $E^{-1.59}$ \citep{bib:090902B_Fermi}, an analysis including size bin 0 and no GH separation cuts would perform better, but the detection would be marginal even if the burst occurred right at zenith. Hence, the analysis starting with bin 1 with GH separation cuts performs better than or equal to the event selection previously used.

Size bin 1 in the Crab analysis has a peak energy of $\sim$550~GeV. For GRB~130427A, the peak energy shifts to $\sim$300 GeV, showing that the
relationship between the size bin number and the energies of the primary particles it includes depends on the gamma-ray spectrum. The difference arises from the fact that EBL absorption removes photons at higher energies. Below $\sim$1~TeV, HAWC can only detect photons if they interact deep in the atmosphere, allowing particles to still reach the detection level. Deeply interacting low-energy photons can have the same event size as a primary gamma ray of higher energy. Since the detector response has been characterized in terms of size bins, the uncertainties derived from the Crab analysis still apply here, even if the energy spectrum is different.

\subsection{Search Circle and Search Duration}
The Crab analysis has demonstrated that the HAWC PSF in size bins 1 to 9 is, within systematic uncertainties, well described by the Monte Carlo simulations and thus the size of the search circle in the current analysis can also be optimised using simulations of gamma-ray air showers at different zenith angles. As the redshift increases, the GRB signal is increasingly dominated by gamma rays of lower energy with worse angular resolution, so the optimal search circle increases. For the exceptional, historic GRBs mentioned above, for GRB~130427A ($z=0.34$), a search circle of $0.8\arcdeg$ is optimal, while for GRB~090510 ($z=0.903$) and GRB~090902B ($z=1.8$), the optimal search circle is $1.0\arcdeg$. For GRB~130427A, a $1.0\arcdeg$ search circle performs only marginally worse ($<10\%$), and since the redshift remains unknown for most GRBs, the angular bin is conservatively chosen to be $1.0\arcdeg$. The optimal search circle does not depend strongly on the zenith angle of the GRB.

In the ON/OFF method, the search circle follows the GRB position for a certain amount of time called the search duration. One search duration used in this paper is the GRB $T_{90}$ derived from observations at lower energies. However, due to the delayed onset and long-lived emission of the high-energy component seen by \emph{Fermi}, this timescale is not necessarily optimal for HAWC. Thus, for long GRBs, the search duration is expanded to $3\times T_{90}$ and $10\times T_{90}$. For short GRBs, even $10\times T_{90}$ might search a very similar time scale to $T_{90}$, so instead 6~s ($=3\times2$~s) and 20~s ($=10\times2$~s) are searched.

\subsection{All-sky Event Rate}
\begin{figure}[t]
\centering
 \plotone{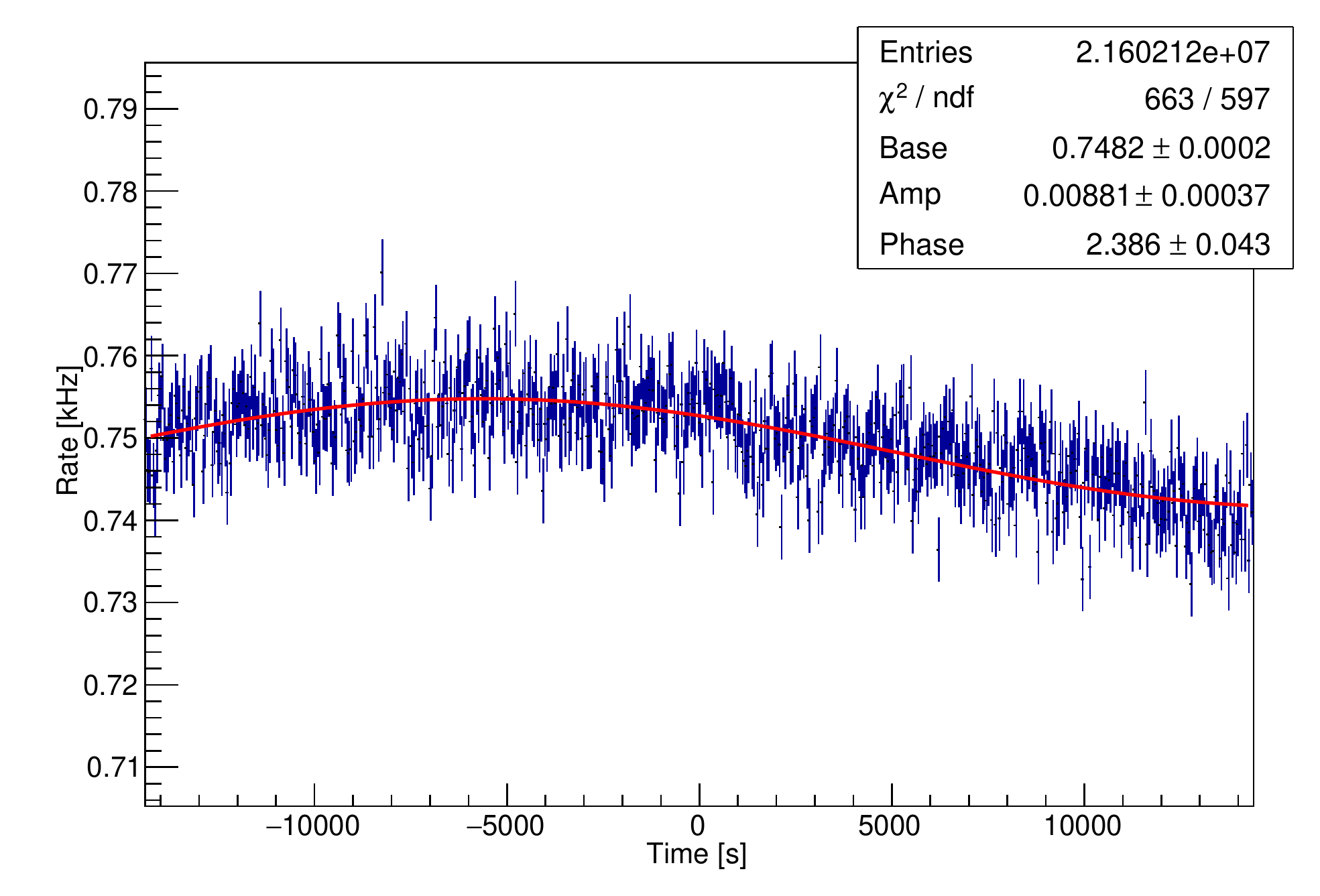}
\caption{All-sky event rate in bins of width $T_{90}=48.0$~s around the trigger time of GRB~151228B at time zero as a representative example. The data are well fitted by a constant (base) plus a sinusoidal modulation. The fitted amplitude is relative to the constant.}
\label{fig:allsky_event_rate_151228B}
\end{figure}

HAWC records data at a rate of about 23~kHz using the current trigger threshold. Restricting the analysis to the single analysis bin discussed earlier (combining size bins 1 to 9 and applying the GH separation cuts for size bin 1) reduces the rate to about 700~Hz. Figure~\ref{fig:allsky_event_rate_151228B} shows the rate around the time of GRB~151205A. The all-sky event rate is well described by a constant plus a sinusoidal oscillation with a 12~hr period. The sinusoidal variation, of the order of $\sim$0.8\%, is caused by atmospheric tides that affect the air shower propagation. The magnitude of this effect is very stable in three years of HAWC data. The sinusoidal is well fit with as little as $\sim$70 minutes of data used, but fails if only one hour of data is used.

The all-sky event rate serves as a tool to assess the data quality. Problems with the detector can result in sudden changes in the rate, which would show up as an unusually high fitted value for the sinusoidal variation. The fit of the all-sky event rate is also used as a way to estimate the expected number of background in the search circle.

\subsection{Background Estimate}
\begin{figure}[t]
\centering
 \plotone{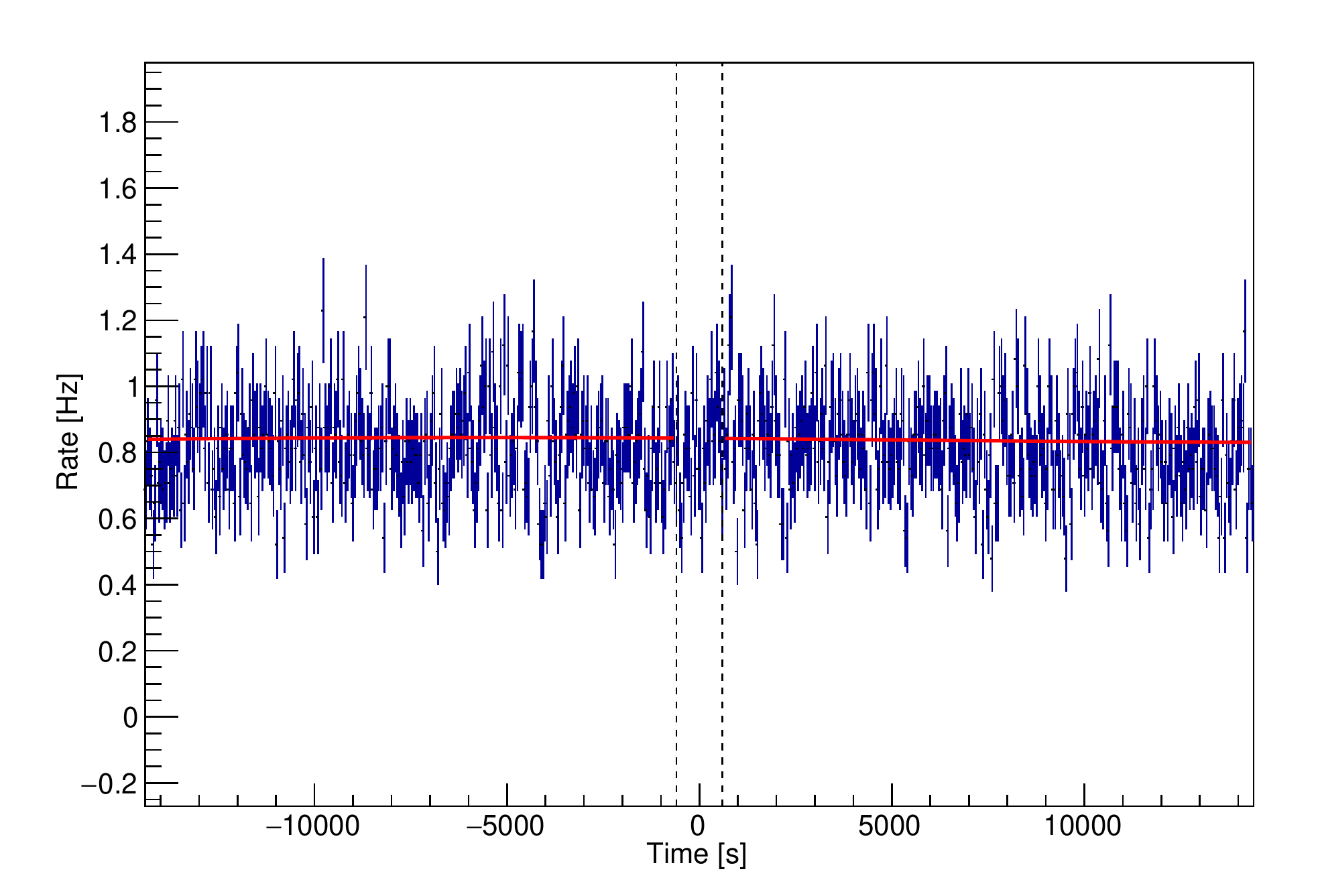}
\caption{Event rate in bins of width $T_{90}=48.0$~s for the on-observation (at zero) and off-observations before and after the trigger time of GRB~151228B (zenith angle $\sim$11$\arcdeg$). The red line shows the background model that is created by scaling the all-sky event rate fit from Fig.~\ref{fig:allsky_event_rate_151228B}. The vertical dashed lines show the region where data are not used for the scaling of the background model.}
\label{fig:on_source_event_rate_151228B}
\end{figure}

The ON/OFF method estimates the background in the search circle using off-observations like the ones shown in Figure~\ref{fig:on_source_event_rate_151228B}. At time zero, the search circle is centred on the GRB position and follows it for the search duration (on-observation). For the other times, the search circle is offset in right ascension by multiples of the search duration (off-observations), thus observing an empty field of the sky that covers the same zenith angles as the on-observation but at a different time. Off-observations are available before and after the GRB, thanks to the continuous data taking of HAWC. For short search durations, the off-observations can spatially overlap. Hence, an off-observation is only taken into account if it has a temporal offset of at least 480~s (corresponding to an angular offset of $\sim$2.0$\arcdeg$) to the previous one. The observations between $\pm$600~s are not used in order to avoid contamination in the off-observations due to the size of the HAWC PSF. The time implies that the closest off-observations have an angular offset of $\sim$2.5$\arcdeg$ from the on-observation, and even at the worst observation conditions, the contamination is less than 5\%.

A background estimate at each point can be obtained using three different methods. In the first method, the off-observations are averaged. The average ignores the sinusoidal variation seen in Fig.~\ref{fig:allsky_event_rate_151228B}, so in the second method, the fit of the all-sky event rate is scaled to match the summed counts in Fig.~\ref{fig:on_source_event_rate_151228B} (again avoiding overlapping off-observations). The resulting line in Fig.~\ref{fig:on_source_event_rate_151228B} is almost indistinguishable from a constant, showing that the sinusoidal oscillation is a negligible correction in a $1\arcdeg$ search circle.

\begin{figure}[t]
\centering
 \plotone{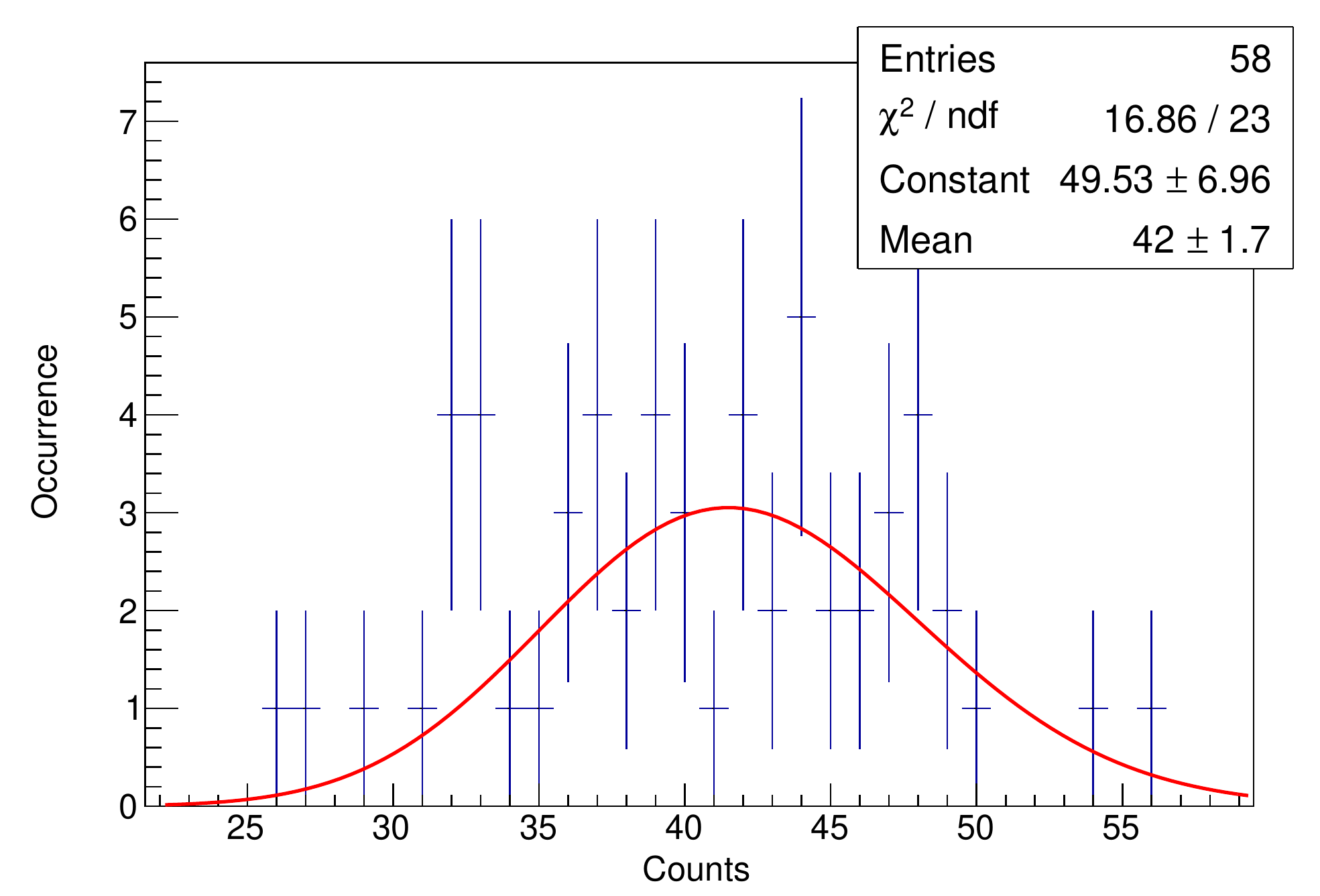}
\caption{Histogram of the counts in the non-overlapping off-observations before and after the trigger time of GRB~151228B from Fig.~\ref{fig:on_source_event_rate_151228B}. The distribution is well fitted by a Poisson distribution with mean $\sim$42.}
\label{fig:poisson_151228B}
\end{figure}

The third method exploits the fact that HAWC is a counting experiment and thus the distribution of counts in the off-observations (again excluding overlapping ones) should follow a Poisson distribution. An independent estimate for the expected background counts in the on-observation can be obtained from the mean of a Poisson distribution fitted to the count distribution (Fig.~\ref{fig:poisson_151228B}). The good fit to a Poisson distribution shows that there is no evidence for correlations between counts that would create a problem for the analysis.

All three methods yield consistent results for all GRBs, short and long, and independent of the burst zenith angle. The successful fit of the Poisson distribution also shows that it is adequate to calculate the significance assuming a Poisson distribution. The default method used is the scaled all-sky event rate. If the all-sky event rate is not available due to insufficient data around the GRB time, the background estimate is obtained using the mean of the off-observations. The background estimate is only deemed reliable if at least 10 independent off-observations are used.

\subsection{Poorly Localised GRBs}
The GRBs detected by the Gamma-ray Burst Monitor on board the \emph{Fermi} satellite \citep[\emph{Fermi}-GBM;][]{bib:GBM_instrument} typically have a positional uncertainty that is wider than the HAWC optimal search circle. In this case, a circle defined by the error radius given by \emph{Fermi}-GBM is covered by multiple search circles, where each circle is offset by 0.3\arcdeg in right ascension or declination until the whole region of interest is covered. The \emph{Fermi}-GBM error radius is the statistical 1$\sigma$ error, but since the errors are not symmetric, the given value is the average of the error ellipse. In the online \emph{Fermi}-GBM Burst Catalog, only the average error radius is available. In addition to the statistical error, \emph{Fermi}-GBM has a systematic uncertainty that is well represented (68\% confidence level) by
a $3.7\arcdeg$ Gaussian with a non-Gaussian tail that contains about 10\% of GBM-detected GRBs and extends to approximately $14\arcdeg$ \citep{bib:GBM_errors}. Limiting the search to the 68\% statistical error implies that occasionally the actual GRB position will not be inside the region of interest. For GRBs close to the edge of the HAWC field of view, it is possible that the error radius extends beyond the HAWC field of view. In those cases, the whole region of interest cannot be covered.

The significance reported here is the highest significance of all search circles. However, this procedure introduces a trial factor. A post-trial significance is derived by using the off-observations. For each off-observation, the most significant result of all search circles is taken and thus a distribution of pre-trial significances is built. The post-trial significance is then obtained by comparing the significance of the on-observation to that distribution. The off-observations are also allowed to overlap in time to increase the statistics of the pre-trial significance distribution.

\section{Results}
During the first one and a half years of HAWC operations with the full HAWC detector, HAWC took data for 72 of 76 GRBs in its field of view, demonstrating the excellent uptime fraction of the detector. The data for nine bursts cannot be analysed due to issues with the data. All of these problems occurred in 2015, while all of the bursts in 2016 can be analysed, since the data taking became more stable. The GRBs analysed and the data used for the analysis are listed in the Appendix (Tables~\ref{tab:LAT_GRBs}--\ref{tab:GBM_GRBs}).

Four GRBs detected by \emph{Fermi}-LAT were inside the HAWC field of view (160503A, 160310A, 150416A and 150314A). Due to power issues, GRB~160503A was missed and thus cannot be analysed. GRB~150416A was only located by \emph{Fermi}-GBM, so it is treated like a GBM burst in the analysis. GRB~150314A, which also triggered \emph{Swift}, occurred at a time when data taking with HAWC was unstable. The analysis would require a different approach, and given the fact that the burst was located at the edge of the HAWC field of view (47\arcdeg), the data were instead inspected by eye and nothing interesting was seen.

In addition to GRB~150314A, 21 \emph{Swift} bursts were inside the HAWC field of view. The detector rate was unstable when GRB~150910A occurred, so these data are not suited for the analysis procedure described earlier. GRB~150323A occurred at a time when data taking with HAWC was unstable, so there is not enough data available to reach 10 independent off-observations. An inspection by eye reveals no outstanding signal, which is not surprising given that these bursts have the two longest $T_{90}$ in the sample and HAWC is not very sensitive to long bursts. The prompt phase of GRB~160624A was not observed by HAWC due to problems with the detector.

During the same time period, \emph{Fermi}-GBM has detected 64 GRBs that were inside the HAWC field of view, which includes the four detections by \emph{Fermi}-LAT mentioned above and nine co-detections with \emph{Swift} (in addition to GRB~150314A), which leaves 51 GRBs observed only by \emph{Fermi}-GBM. For bn~160609690 and bn~150520893, no HAWC data are available. The data for bn~151211672, bn~150911315, bn~150819440, bn~150612702, bn~150528656 and bn~150208929 show instability in the trigger rate, and these bursts are thus not analysed.

GRB~170206A is the third brightest short GRB detected by \emph{Fermi}-GBM so far and occurred during the writing of this paper. The low zenith angle of 11\arcdeg and the fact that the burst was also detected by \emph{Fermi}-LAT \citep{bib:170206A_LAT} make this burst an excellent candidate for HAWC. The burst is included in this paper using data reconstructed directly at the HAWC site. Differences from the current offline analysis are only minor.

Table~\ref{tab:Swift_results_T90} and Table~\ref{tab:GBM_results_T90} show the results for the analysis using the low energy $T_{90}$. Table~\ref{tab:Swift_results_3T90}--\ref{tab:GBM_results_10T90} show the results for the extended time periods. All GRBs are consistent with the assumption of background only. Event upper limits are derived as the upper edge of frequentist confidence intervals \citep{bib:pdg}. A likelihood-ratio ordering principle first described by \cite{bib:Feldman_Cousins} is used. The upper limits are converted to a flux limit using the HAWC effective area derived from the Monte Carlo described in \cite{bib:HAWC_crab}. Flux limits are given for a power-law spectrum with an index of $-2.0$, which is the average value for the photon index of the high-energy component measured by \emph{Fermi}-LAT \citep{bib:Fermi_LAT_GRB_catalogue}, and assuming EBL absorption according to the fiducial model in \cite{bib:EBL}. Short GRBs might have a somewhat harder index, as seen in GRB~081024B \citep{bib:081024B_Fermi} and GRB~090510 (see above). Assuming a photon index of $-1.7$ would improve the limits by a factor of about 6 for a burst at a redshift of 0.3 at zenith and 7 for a burst at a zenith angle of 40\arcdeg. For a burst at a redshift of 1.0, the limits would improve by a factor of 4.

The energy range of these limits, described by the energy range that gives the central 90\% of the total number of expected counts, depends on the zenith angle and redshift of the burst. For a burst with a power-law spectrum with index $-2$ at a redshift of 0.3, the range is about 100~GeV to 1~TeV at zenith and increases to about 150~GeV to 3~TeV at a zenith angle of 40\arcdeg. If the burst is at a redshift of 1.0, EBL absorption removes most of the expected counts at higher energies, making the 90\% range very narrow. The lower bound changes from 50 to 70~GeV when going from zenith to a zenith angle of 40\arcdeg, while the upper end of the 90\% range ends at about 200~GeV. This shows again that the relationship between the size bin number and the energies of the primary particles it includes depends on the gamma-ray spectrum.

\begin{deluxetable*}{l|cccccccc}
\tabletypesize{\footnotesize}
\tablecolumns{9}
\tablewidth{0pt}
\tablecaption{Results for \emph{Fermi}-LAT- and \emph{Swift}-detected GRBs \label{tab:Swift_results_T90}}
\tablehead{\vspace{-0.1cm} & \colhead{Measure-}   & \colhead{Background} & \colhead{Signif-} & \colhead{Upper} & \colhead{Sensi-}
                           & \colhead{$A$ at 1~GeV\tablenotemark{a}} & \colhead{$A$ at 1~GeV\tablenotemark{a}}
                           & \colhead{$A$ at 1~GeV\tablenotemark{a}}\\
           \vspace{-0.1cm} & \colhead{ment}       & \colhead{Estimate}   & \colhead{icance}  & \colhead{Limit} & \colhead{tivity}
                           & \colhead{$z=0.3$}                     & \colhead{$z=1.0$}
                           & \colhead{GRB $z$}}
\startdata
170206A & 0 & 0.90 & ... & 1.67 & 3.19 & 6.05e--11 & 1.20e--09 & ...\\
160310A & 5 & 4.46 & 0.10 & 5.53 & 4.98 & 1.92e--10 & 6.26e--09 & ...\\
\hline
160410A & 2 & 2.49 & -0.55 & 3.46 & 4.19 & --- & --- & 3.88e--08\\
151228B & 36 & 40.45 & -0.77 & 7.10 & 12.22 & 6.28e--12 & 1.25e--10 & ...\\
151205A & 36 & 32.58 & 0.53 & 14.45 & 11.11 & 1.95e--11 & 4.26e--10 & ...\\
150817A & 7 & 11.37 & -1.52 & 2.55 & 7.06 & 2.97e--11 & 9.60e--10 & ...\\
150811A & 2 & 6.41 & -2.25 & 1.36 & 5.66 & 2.53e--11 & 8.27e--10 & ...\\
150716A & 5 & 5.94 & -0.54 & 4.12 & 5.51 & 2.32e--10 & 1.02e--08 & ...\\
150710A & 0 & 0.10 & ... & 2.33 & 2.52 & 5.49e--10 & 8.52e--09 & ...\\
150530B & 10 & 7.06 & 0.93 & 9.44 & 5.90 & 1.03e--10 & 2.91e--09 & ...\\
150530A & 1 & 1.18 & -0.50 & 3.19 & 3.40 & 7.19e--10 & 2.41e--08 & ...\\
150527A & 20 & 12.74 & 1.80 & 15.75 & 7.41 & 3.49e--10 & 1.54e--08 & ...\\
150428A & 1 & 2.60 & -1.44 & 2.12 & 4.24 & 8.12e--10 & 1.97e--08 & ...\\
150423A & 0 & 0.12 & ... & 2.31 & 2.54 & 5.09e--10 & 9.33e--09 & 2.85e--08\\
150323B & 9 & 11.96 & -1.00 & 3.95 & 7.23 & 6.50e--11 & 1.57e--09 & ...\\
150317A & 4 & 4.29 & -0.31 & 4.30 & 4.92 & 1.82e--10 & 4.40e--09 & ...\\
150302A & 13 & 8.39 & 1.38 & 11.67 & 6.29 & 1.52e--10 & 4.12e--09 & ...\\
150211A & 1 & 1.09 & -0.42 & 3.27 & 3.34 & 1.90e--09 & 7.38e--08 & ...\\
150203A & 15 & 21.88 & -1.64 & 2.75 & 9.35 & 5.16e--12 & 9.45e--11 & ...\\
141205A & 2 & 0.74 & 0.95 & 5.17 & 3.06 & 3.38e--10 & 7.64e--09 & ...\\
\enddata
\tablecomments{Results for \emph{Fermi}-LAT-detected bursts 170206A and 160310A and the 18 \emph{Swift}-detected GRBs in the HAWC field of view that are analysed in this paper. The \emph{Fermi}-LAT-detected burst GRB~150416A can be found in Table~\ref{tab:GBM_results_T90}. For each GRB, the measurement and the corresponding background estimate from the data are given (in units of counts). The corresponding $p$-value from a Poisson distribution is converted to the standard deviation of a Gaussian distribution (significance), except when the measurement is zero, which means the $p$-value is 1, and an equivalent standard deviation is undefined. The event upper limits correspond to a 90\% confidence level and sensitivity is the corresponding average event upper limit as defined in \cite{bib:Feldman_Cousins}. Flux limits are given for a power law $A\left(\frac{E}{1~\rm{GeV}}\right)^{-2.0}$ assuming redshifts of 0.3, 1.0 or the GRB redshift if known. See the main text for the validity range of the limits. The redshift for 150423A is only tentative, so all three options are given.}
\tablenotetext{a}{($\rm{cm}^{-2}\rm{s}^{-1}\rm{keV}^{-1}$)}
\end{deluxetable*}

\begin{deluxetable*}{l|cccccccccc}
\tabletypesize{\footnotesize}
\tablecolumns{9}
\tablewidth{0pt}
\tablecaption{Results for \emph{Fermi}-GBM-detected GRBs \label{tab:GBM_results_T90}}
\tablehead{\vspace{-0.1cm} & \colhead{Measure-}   & \colhead{Background} & \colhead{Signif-} & \colhead{Post-Trial}
                           & \colhead{R.A.} & \colhead{Decl.} & \colhead{Upper} & \colhead{Sensi-}
                           & \colhead{$A$ at 1~GeV\tablenotemark{a}} & \colhead{$A$ at 1~GeV\tablenotemark{a}}\\
           \vspace{-0.1cm} & \colhead{ment}       & \colhead{Estimate}   & \colhead{icance}  & \colhead{Signif.}
                           & \colhead{(deg)} & \colhead{(deg)} & \colhead{Limit} & \colhead{tivity}
                           & \colhead{$z=0.3$} & \colhead{$z=1.0$}}
\startdata
bn160605847 & 4 & 0.97 & 2.11 & -0.51 & 107.37 & -18.1 & 7.62 & 3.25 & 1.37e--09 & 3.29e--08\\
bn160527080 & 9 & 3.70 & 2.21 & -1.46 & 216.50 & 10.5 & 11.60 & 4.70 & 6.84e--10 & 2.29e--08\\
bn160521839 & 10 & 3.72 & 2.57 & 0.12 & 75.79 & -14.0 & 12.78 & 4.71 & 5.08e--10 & 1.66e--08\\
bn160515819 & 22 & 14.47 & 1.76 & -1.40 & 84.18 & -16.0 & 16.50 & 7.82 & 1.93e--10 & 4.64e--09\\
bn160406503 & 3 & 0.26 & 2.81 & -0.05 & 259.06 & 21.2 & 7.16 & 2.66 & 1.19e--09 & 2.70e--08\\
*bn160315739 & 3 & 0.21 & 3.01 & -0.59 & 305.29 & -8.2 & 7.21 & 2.62 & 1.72e--08 & 6.65e--07\\
*bn160303201 & 9 & 1.92 & 3.58 & 1.12 & 161.59 & 65.3 & 13.38 & 3.87 & 1.02e--08 & 1.34e--06\\
bn160301215 & 16 & 9.29 & 1.91 & -0.92 & 113.11 & 2.3 & 14.71 & 6.56 & 1.53e--10 & 4.15e--09\\
*bn160228034 & 12 & 3.71 & 3.30 & 0.29 & 43.23 & 43.7 & 15.29 & 4.71 & 5.98e--10 & 1.95e--08\\
*bn160220868 & 23 & 10.69 & 3.18 & -0.47 & 340.67 & 13.6 & 21.30 & 6.92 & 1.25e--10 & 3.13e--09\\
bn160215773 & 54 & 37.82 & 2.42 & -0.12 & 354.60 & 0.0 & 29.65 & 11.85 & 9.47e--11 & 3.07e--09\\
bn160211119 & 2 & 0.08 & 2.71 & 1.19 & 117.06 & 56.1 & 5.83 & 2.51 & 2.85e--08 & 8.83e--07\\
bn160206430 & 11 & 4.32 & 2.58 & 0.17 & 181.41 & 53.8 & 13.48 & 4.93 & 6.19e--10 & 1.49e--08\\
bn160131174 & 165 & 132.24 & 2.72 & -0.28 & 112.64 & 15.2 & 55.25 & 21.13 & 1.69e--11 & 3.56e--10\\
bn160118060 & 7 & 3.91 & 1.27 & -0.37 & 19.17 & 59.9 & 8.62 & 4.78 & 8.64e--10 & 2.68e--08\\
bn160113398 & 13 & 9.33 & 1.04 & -0.52 & 186.54 & 12.0 & 10.72 & 6.57 & 9.51e--11 & 2.69e--09\\
bn160107931 & 11 & 7.77 & 0.99 & ... & ... & ... & 10.04 & 6.12 & 6.98e--10 & 2.70e--08\\
bn160102500 & 11 & 3.96 & 2.79 & 0.22 & 223.36 & 11.4 & 13.85 & 4.80 & 8.16e--10 & 2.73e--08\\
bn151129333 & 11 & 3.75 & 2.92 & 0.41 & 57.93 & -12.6 & 14.06 & 4.72 & 2.13e--09 & 8.26e--08\\
bn151030999 & 91 & 86.27 & 0.47 & ... & ... & ... & 21.75 & 17.33 & 9.03e--12 & 1.65e--10\\
*bn151023104 & 8 & 2.06 & 3.01 & -0.43 & 349.84 & -11.7 & 11.93 & 3.94 & 7.35e--10 & 2.40e--08\\
*bn151022577 & 2 & 0.10 & 2.61 & -0.94 & 108.39 & 21.3 & 5.81 & 2.52 & 5.62e--09 & 1.52e--07\\
bn150928359 & 10 & 2.96 & 3.09 & 0.97 & 89.52 & 34.8 & 13.54 & 4.39 & 5.15e--09 & 1.25e--07\\
*bn150923297 & 1 & 0.02 & 2.09 & -0.01 & 325.21 & 24.7 & 4.33 & 2.45 & 1.06e--07 & 3.29e--06\\
bn150906944 & 2 & 0.15 & 2.31 & 0.32 & 215.84 & -1.7 & 5.75 & 2.57 & 1.83e--09 & 3.97e--08\\
*bn150904479 & 4 & 0.73 & 2.48 & -2.00 & 74.39 & -28.1 & 7.87 & 3.05 & 1.24e--08 & 1.63e--06\\
bn150811849 & 0 & 0.09 & ... & ... & ... & ... & 2.34 & 2.52 & 3.61e--09 & 8.70e--08\\
bn150721242 & 3 & 1.36 & 1.00 & -0.51 & 333.12 & 7.5 & 6.06 & 3.53 & 2.60e--09 & 1.01e--07\\
bn150705588 & 2 & 0.16 & 2.26 & -0.97 & 64.33 & -3.6 & 5.75 & 2.58 & 5.15e--09 & 1.68e--07\\
bn150622393 & 2 & 4.12 & -1.38 & ... & ... & ... & 2.25 & 4.86 & 2.94e--10 & 1.14e--08\\
bn150522944 & 3 & 0.24 & 2.91 & 0.54 & 121.03 & 53.7 & 7.19 & 2.64 & 4.43e--09 & 1.45e--07\\
bn150502435 & 15 & 15.93 & -0.32 & ... & ... & ... & 6.60 & 8.14 & 9.02e--11 & 3.02e--09\\
bn150416773 & 8 & 3.28 & 2.07 & 0.48 & 60.98 & 51.8 & 10.71 & 4.53 & 1.51e--09 & 4.68e--08\\
*bn150329288 & 8 & 1.58 & 3.49 & 0.92 & 170.92 & -21.4 & 12.41 & 3.67 & 3.40e--09 & 1.32e--07\\
*bn150318521 & 4 & 2.05 & 1.03 & -1.39 & 271.40 & -31.1 & 6.54 & 3.94 & 7.26e--09 & 4.96e--07\\
*bn150206407 & 5 & 0.55 & 3.46 & 1.73 & 212.41 & 57.1 & 9.44 & 2.89 & 8.66e--09 & 2.68e--07\\
*bn150201040 & 2 & 0.13 & 2.44 & -0.57 & 11.13 & 15.9 & 5.78 & 2.55 & 7.12e--09 & 2.33e--07\\
bn150131951 & 7 & 1.41 & 3.22 & 1.44 & 57.64 & 22.0 & 11.12 & 3.56 & 2.03e--09 & 6.79e--08\\
bn150128791 & 19 & 8.38 & 3.06 & 1.12 & 269.04 & 27.5 & 19.12 & 6.28 & 1.05e--09 & 3.27e--08\\
bn150126868 & 31 & 28.53 & 0.40 & ... & ... & ... & 12.95 & 10.46 & 6.06e--11 & 1.96e--09\\
bn150110433 & 4 & 5.33 & -0.77 & ... & ... & ... & 3.33 & 5.29 & 9.10e--10 & 2.21e--08\\
bn150105257 & 9 & 8.98 & -0.10 & ... & ... & ... & 6.32 & 6.47 & 2.13e--10 & 9.35e--09\\
bn141230142 & 14 & 6.34 & 2.53 & 0.15 & 58.98 & 0.7 & 15.16 & 5.63 & 1.11e--10 & 2.50e--09\\
bn141202470 & 1 & 0.12 & 1.23 & 0.04 & 145.94 & 63.1 & 4.24 & 2.54 & 2.50e--08 & 9.68e--07\\
\enddata
\tablecomments{Same as Table~\ref{tab:Swift_results_T90}, but for the 44 \emph{Fermi}-GBM-detected GRBs in the HAWC field of view that are analysed in this paper. An asterisk indicates a GRB with an error radius that extends beyond the HAWC field of view. An additional column shows the significance after accounting for trials and the right ascension / declination of the location of the tile. GRBs with an error radius smaller than the HAWC search circle size have no post-trial significance and are centered on the GRB location.}
\tablenotetext{a}{($\rm{cm}^{-2}\rm{s}^{-1}\rm{keV}^{-1}$)}
\end{deluxetable*}

\begin{deluxetable*}{l|cccccccc}
\tabletypesize{\footnotesize}
\tablecolumns{9}
\tablewidth{0pt}
\tablecaption{Results for \emph{Fermi}- and \emph{Swift}-detected GRBs (Extended Time Search I) \label{tab:Swift_results_3T90}}
\tablehead{\vspace{-0.1cm} & \colhead{Measure-}   & \colhead{Background} & \colhead{Signif-} & \colhead{Upper} & \colhead{Sensi-}
                           & \colhead{$A$ at 1~GeV\tablenotemark{a}} & \colhead{$A$ at 1~GeV\tablenotemark{a}}
                           & \colhead{$A$ at 1~GeV\tablenotemark{a}}\\
           \vspace{-0.1cm} & \colhead{ment}       & \colhead{Estimate}   & \colhead{icance}  & \colhead{Limit} & \colhead{tivity}
                           & \colhead{$z=0.3$}                     & \colhead{$z=1.0$}
                           & \colhead{GRB $z$}}
\startdata
170206A & 5 & 3.82 & 0.42 & 6.17 & 4.75 & 4.36e--11 & 8.66e--10 & ...\\
160310A & 16 & 13.87 & 0.47 & 10.10 & 7.68 & 1.17e--10 & 3.82e--09 & ...\\
\hline
160410A & 9 & 6.84 & 0.67 & 8.46 & 5.83 & ... & ... & 3.16e--08\\
151228B & 102 & 117.05 & -1.46 & 6.70 & 19.96 & 1.97e--12 & 3.92e--11 & ...\\
151205A & 107 & 95.83 & 1.09 & 29.20 & 18.16 & 1.32e--11 & 2.87e--10 & ...\\
150817A & 27 & 33.69 & -1.26 & 4.55 & 11.25 & 1.76e--11 & 5.71e--10 & ...\\
150811A & 12 & 22.40 & -2.50 & 1.80 & 9.45 & 1.11e--11 & 3.63e--10 & ...\\
150716A & 20 & 17.46 & 0.52 & 11.05 & 8.49 & 2.08e--10 & 9.14e--09 & ...\\
150710A & 8 & 6.25 & 0.55 & 7.74 & 5.61 & 4.55e--11 & 7.05e--10 & ...\\
150530B & 29 & 23.04 & 1.13 & 15.95 & 9.59 & 5.79e--11 & 1.64e--09 & ...\\
150530A & 2 & 3.32 & -1.01 & 2.79 & 4.54 & 2.10e--10 & 7.04e--09 & ...\\
150527A & 47 & 40.72 & 0.91 & 18.80 & 12.25 & 1.39e--10 & 6.11e--09 & ...\\
150428A & 7 & 7.89 & -0.45 & 4.68 & 6.15 & 5.96e--10 & 1.45e--08 & ...\\
150423A & 1 & 4.87 & -2.42 & 1.26 & 5.13 & 1.01e--11 & 1.85e--10 & 5.67e--10\\
150323B & 35 & 36.40 & -0.29 & 9.65 & 11.67 & 5.29e--11 & 1.27e--09 & ...\\
150317A & 11 & 13.43 & -0.78 & 4.35 & 7.55 & 6.14e--11 & 1.48e--09 & ...\\
150302A & 34 & 25.29 & 1.59 & 19.70 & 9.94 & 8.56e--11 & 2.32e--09 & ...\\
150211A & 7 & 3.61 & 1.45 & 8.92 & 4.67 & 1.73e--09 & 6.71e--08 & ...\\
150203A & 62 & 67.25 & -0.69 & 9.30 & 15.40 & 5.81e--12 & 1.06e--10 & ...\\
141205A & 6 & 4.09 & 0.74 & 7.38 & 4.85 & 8.84e--11 & 2.00e--09 & ...\\
\enddata
\tablecomments{Same as Table~\ref{tab:Swift_results_T90}, but using a search duration of three times $T_{90}$ for long GRBs and six~s for short GRBs.}
\tablenotetext{a}{($\rm{cm}^{-2}\rm{s}^{-1}\rm{keV}^{-1}$)}
\end{deluxetable*}

\begin{deluxetable*}{l|cccccccc}
\tabletypesize{\footnotesize}
\tablecolumns{9}
\tablewidth{0pt}
\tablecaption{Results for \emph{Fermi}- and \emph{Swift}-detected GRBs (Extended Time Search II) \label{tab:Swift_results_10T90}}
\tablehead{\vspace{-0.1cm} & \colhead{Measure-}   & \colhead{Background} & \colhead{Signif-} & \colhead{Upper} & \colhead{Sensi-}
                           & \colhead{$A$ at 1~GeV\tablenotemark{a}} & \colhead{$A$ at 1~GeV\tablenotemark{a}}
                           & \colhead{$A$ at 1~GeV\tablenotemark{a}}\\
           \vspace{-0.1cm} & \colhead{ment}       & \colhead{Estimate}   & \colhead{icance}  & \colhead{Limit} & \colhead{tivity}
                           & \colhead{$z=0.3$}                     & \colhead{$z=1.0$}
                           & \colhead{GRB $z$}}
\startdata
170206A & 12 & 13.77 & -0.58 & 5.35 & 7.68  & 1.13e--11 & 2.25e--10 & ...\\
160310A & 53 & 45.82 & 0.99 & 20.20 & 12.94 & 7.01e--11 & 2.29e--09 & ...\\
\hline
160410A & 27 & 23.84 & 0.57 & 13.15 & 9.71 & ... & ... & 1.48e--08\\
151228B & 392 & 389.71 & 0.10 & 35.50 & 35.12 & 3.14e--12 & 6.23e--11 & ...\\
151205A & 338 & 333.61 & 0.22 & 35.50 & 32.53 & 4.80e--12 & 1.05e--10 & ...\\
150817A & 114 & 107.06 & 0.63 & 25.45 & 19.12 & 2.96e--11 & 9.59e--10 & ...\\
150811A & 71 & 76.80 & -0.71 & 9.85 & 16.38 & 1.83e--11 & 5.97e--10 & ...\\
150710A & 16 & 17.76 & -0.51 & 6.35 & 8.56 & 1.12e--11 & 1.74e--10 & ...\\
150530B & 83 & 77.07 & 0.63 & 21.95 & 16.42 & 2.39e--11 & 6.77e--10 & ...\\
150530A & 10 & 10.11 & -0.14 & 6.35 & 6.73 & 1.43e--10 & 4.80e--09 & ...\\
150527A & 154 & 155.90 & -0.18 & 19.50 & 22.66 & 4.32e--11 & 1.90e--09 & ...\\
150428A & 20 & 28.26 & -1.71 & 2.80 & 10.42 & 1.07e--10 & 2.60e--09 & ...\\
150423A & 9 & 15.21 & -1.83 & 2.35 & 8.01 & 5.68e--12 & 1.04e--10 & 3.19e--10\\
150323B & 112 & 129.03 & -1.57 & 6.35 & 20.90 & 1.04e--11 & 2.52e--10 & ...\\
150317A & 31 & 42.01 & -1.84 & 3.30 & 12.42 & 1.40e--11 & 3.37e--10 & ...\\
150302A & 91 & 85.05 & 0.60 & 22.95 & 17.19 & 2.99e--11 & 8.10e--10 & ...\\
150211A & 21 & 12.72 & 2.05 & 17.25 & 7.40 & 1.00e--09 & 3.89e--08 & ...\\
150203A & 195 & 221.34 & -1.83 & 6.50 & 26.78 & 1.22e--12 & 2.23e--11 & ...\\
141205A & 12 & 13.14 & -0.42 & 5.90 & 7.51 & 2.12e--11 & 4.80e--10 & ...\\
\enddata
\tablecomments{Same as Table~\ref{tab:Swift_results_T90}, but using a search duration of 10 times $T_{90}$ for long GRBs and 20~s for short GRBs. There is not enough data for the analysis of GRB~150716A for this search duration.}
\tablenotetext{a}{($\rm{cm}^{-2}\rm{s}^{-1}\rm{keV}^{-1}$)}
\end{deluxetable*}

\begin{deluxetable*}{l|cccccccccc}
\tabletypesize{\footnotesize}
\tablecolumns{9}
\tablewidth{0pt}
\tablecaption{Results for \emph{Fermi}-GBM-detected GRBs (Extended Time Search I) \label{tab:GBM_results_3T90}}
\tablehead{\vspace{-0.1cm} & \colhead{Measure-}   & \colhead{Background} & \colhead{Signif-} & \colhead{Post-Trial}                            
                           & \colhead{R.A.} & \colhead{Decl.} & \colhead{Upper} & \colhead{Sensi-}
                           & \colhead{$A$ at 1~GeV\tablenotemark{a}} & \colhead{$A$ at 1~GeV\tablenotemark{a}}\\
           \vspace{-0.1cm} & \colhead{ment}       & \colhead{Estimate}   & \colhead{icance}  & \colhead{Signif.} 
                           & \colhead{(deg)} & \colhead{(deg)} & \colhead{Limit} & \colhead{tivity} 
                           & \colhead{$z=0.3$} & \colhead{$z=1.0$}}
\startdata
bn160605847 & 8 & 3.35 & 2.03 & -1.18 & 108.34 & -16.6 & 10.64 & 4.56 & 6.36e--10 & 1.53e--08\\
bn160527080 & 20 & 9.50 & 2.88 & 0.05 & 214.43 & 1.8 & 19.02 & 6.61 & 6.21e--10 & 2.73e--08\\
bn160521839 & 25 & 12.70 & 2.97 & 0.68 & 73.62 & -11.9 & 21.80 & 7.40 & 2.07e--10 & 6.69e--09\\
bn160515819 & 55 & 40.48 & 2.12 & -0.46 & 82.64 & -17.2 & 28.00 & 12.22 & 1.09e--10 & 2.63e--09\\
bn160406503 & 13 & 3.92 & 3.51 & 0.75 & 274.48 & 35.6 & 16.13 & 4.79 & 1.68e--10 & 3.56e--09\\
*bn160315739 & 5 & 0.50 & 3.58 & 0.60 & 304.04 & -9.1 & 9.48 & 2.85 & 1.93e--08 & 4.69e--07\\
*bn160303201 & 23 & 9.64 & 3.57 & 0.85 & 153.10 & 63.2 & 22.27 & 6.65 & 1.22e--09 & 4.73e--08\\
bn160301215 & 42 & 28.97 & 2.21 & -0.43 & 114.91 & 1.1 & 25.00 & 10.57 & 8.68e--11 & 2.35e--09\\
*bn160228034 & 24 & 11.09 & 3.28 & 0.06 & 43.23 & 43.7 & 22.40 & 7.02 & 2.92e--10 & 9.54e--09\\
*bn160220868 & 39 & 20.71 & 3.52 & 0.22 & 336.44 & 6.1 & 29.80 & 9.13 & 1.36e--10 & 3.70e--09\\
bn160215773 & 112 & 90.77 & 2.12 & -0.95 & 360.00 & 2.7 & 39.75 & 17.72 & 9.25e--11 & 2.23e--09\\
bn160211119 & 3 & 0.55 & 2.10 & -0.26 & 117.11 & 55.8 & 6.88 & 2.89 & 5.38e--09 & 1.67e--07\\
bn160206430 & 22 & 11.98 & 2.52 & -0.09 & 180.85 & 54.4 & 19.00 & 7.24 & 2.91e--10 & 7.00e--09\\
bn160118060 & 17 & 12.03 & 1.27 & -0.51 & 18.56 & 59.3 & 12.95 & 7.25 & 4.33e--10 & 1.34e--08\\
bn160113398 & 35 & 28.21 & 1.18 & -0.42 & 186.54 & 12.0 & 17.80 & 10.41 & 5.26e--11 & 1.49e--09\\
bn160107931 & 27 & 21.21 & 1.14 & ... & ... & ... & 15.75 & 9.21 & 3.65e--10 & 1.41e--08\\
bn160102500 & 26 & 13.52 & 2.94 & 0.28 & 221.83 & 10.5 & 22.00 & 7.58 & 2.86e--10 & 6.88e--09\\
bn151129333 & 22 & 11.15 & 2.79 & 0.05 & 57.93 & -12.6 & 19.85 & 7.02 & 1.00e--09 & 3.89e--08\\
bn151030999 & 277 & 264.04 & 0.77 & ... & ... & ... & 41.50 & 29.19 & 5.75e--12 & 1.05e--10\\
*bn151023104 & 15 & 5.73 & 3.12 & -0.45 & 350.42 & -12.6 & 16.79 & 5.43 & 5.39e--10 & 1.30e--08\\
*bn151022577 & 7 & 1.31 & 3.34 & -0.34 & 107.20 & 38.4 & 11.21 & 3.49 & 1.18e--09 & 3.85e--08\\
bn150928359 & 22 & 9.62 & 3.34 & 1.24 & 88.76 & 34.2 & 21.36 & 6.64 & 2.71e--09 & 6.56e--08\\
*bn150923297 & 4 & 0.86 & 2.27 & -0.69 & 329.09 & 30.1 & 7.74 & 3.16 & 1.93e--09 & 6.45e--08\\
bn150906944 & 11 & 3.77 & 2.91 & 0.61 & 214.35 & 5.5 & 14.04 & 4.73 & 1.68e--10 & 3.81e--09\\
*bn150904479 & 12 & 4.73 & 2.69 & -1.59 & 70.55 & -24.5 & 14.27 & 5.09 & 1.62e--09 & 6.26e--08\\
bn150811849 & 0 & 0.97 & ... & ... & ... & ... & 1.62 & 3.25 & 2.66e--10 & 6.42e--09\\
bn150721242 & 5 & 3.89 & 0.39 & -1.80 & 333.73 & 7.5 & 6.10 & 4.77 & 8.74e--10 & 3.38e--08\\
bn150705588 & 8 & 2.33 & 2.77 & -0.95 & 59.23 & 1.5 & 11.65 & 4.12 & 3.28e--10 & 7.90e--09\\
bn150622393 & 10 & 12.53 & -0.85 & ... & ... & ... & 4.25 & 7.36 & 1.85e--10 & 7.16e--09\\
bn150522944 & 7 & 1.13 & 3.57 & 1.28 & 139.71 & 53.4 & 11.39 & 3.37 & 1.87e--09 & 4.52e--08\\
bn150502435 & 45 & 44.57 & 0.02 & ... & ... & ... & 12.50 & 12.77 & 5.69e--11 & 1.91e--09\\
bn150416773 & 20 & 10.11 & 2.66 & 0.97 & 59.54 & 52.4 & 18.40 & 6.73 & 8.66e--10 & 2.68e--08\\
*bn150329288 & 24 & 11.64 & 3.09 & -0.31 & 166.24 & -10.0 & 21.85 & 7.14 & 3.76e--10 & 1.26e--08\\
*bn150318521 & 9 & 6.24 & 0.92 & -1.61 & 271.40 & -31.1 & 9.05 & 5.61 & 3.35e--09 & 2.29e--07\\
*bn150206407 & 10 & 2.10 & 3.81 & 2.14 & 208.73 & 56.5 & 14.39 & 3.97 & 2.33e--09 & 1.02e--07\\
*bn150201040 & 7 & 1.61 & 2.99 & -0.45 & 11.78 & 16.8 & 10.92 & 3.69 & 1.15e--09 & 3.75e--08\\
bn150131951 & 10 & 3.20 & 2.91 & 0.67 & 60.52 & 24.1 & 13.29 & 4.50 & 1.34e--09 & 5.91e--08\\
bn150128791 & 66 & 41.33 & 3.49 & 2.36 & 273.76 & 26.6 & 39.20 & 12.32 & 2.29e--10 & 7.67e--09\\
bn150126868 & 90 & 87.85 & 0.19 & ... & ... & ... & 19.15 & 17.44 & 2.99e--11 & 9.67e--10\\
bn150110433 & 19 & 17.65 & 0.24 & ... & ... & ... & 9.85 & 8.55 & 8.99e--10 & 2.18e--08\\
bn150105257 & 24 & 24.72 & -0.21 & ... & ... & ... & 8.75 & 9.85 & 9.82e--11 & 4.32e--09\\
bn141230142 & 36 & 20.10 & 3.13 & 1.04 & 53.88 & 0.7 & 26.90 & 9.01 & 6.54e--11 & 1.48e--09\\
bn141202470 & 3 & 0.77 & 1.72 & -0.57 & 142.28 & 60.1 & 6.66 & 3.08 & 2.75e--09 & 1.21e--07\\
\enddata
\tablecomments{Same as Table~\ref{tab:GBM_results_T90}, but using a search duration of three times $T_{90}$ for long GRBs and six~s for short GRBs. There is not enough data for the analysis of bn160131174 for this search duration.}
\tablenotetext{a}{($\rm{cm}^{-2}\rm{s}^{-1}\rm{keV}^{-1}$)}
\end{deluxetable*}

\begin{deluxetable*}{l|cccccccccc}
\tabletypesize{\footnotesize}
\tablecolumns{9}
\tablewidth{0pt}
\tablecaption{Results for \emph{Fermi}-GBM-detected GRBs (Extended Time Search II)\label{tab:GBM_results_10T90}}
\tablehead{\vspace{-0.1cm} & \colhead{Measure-}   & \colhead{Background} & \colhead{Signif-} & \colhead{Post-Trial}
                           & \colhead{R.A.} & \colhead{Decl.} & \colhead{Upper} & \colhead{Sensi-}
                           & \colhead{$A$ at 1~GeV\tablenotemark{a}} & \colhead{$A$ at 1~GeV\tablenotemark{a}}\\
           \vspace{-0.1cm} & \colhead{ment}       & \colhead{Estimate}   & \colhead{icance}  & \colhead{Signif.}
                           & \colhead{(deg)} & \colhead{(deg)} & \colhead{Limit} & \colhead{tivity} 
                           & \colhead{$z=0.3$} & \colhead{$z=1.0$}}
\startdata
bn160605847 & 25 & 11.22 & 3.46 & 1.72 & 108.34 & -16.6 & 23.25 & 7.03 & 4.17e--10 & 1.00e--08\\
bn160527080 & 57 & 37.29 & 2.95 & -0.09 & 213.82 & 3.3 & 33.20 & 11.77 & 1.96e--10 & 6.55e--09\\
bn160521839 & 42 & 23.36 & 3.41 & 1.46 & 72.07 & -17.6 & 30.65 & 9.63 & 2.88e--10 & 9.66e--09\\
bn160406503 & 27 & 13.45 & 3.18 & -0.33 & 274.02 & 35.0 & 23.55 & 7.57 & 7.38e--11 & 1.56e--09\\
*bn160315739 & 27 & 11.61 & 3.78 & 0.86 & 331.34 & -7.9 & 25.35 & 7.16 & 1.66e--10 & 4.70e--09\\
*bn160303201 & 57 & 32.58 & 3.82 & 1.50 & 153.77 & 63.2 & 37.95 & 11.10 & 6.25e--10 & 2.42e--08\\
bn160301215 & 134 & 105.15 & 2.67 & 0.43 & 117.01 & 1.4 & 48.85 & 18.98 & 3.59e--11 & 1.01e--09\\
*bn160228034 & 91 & 58.18 & 3.93 & 1.36 & 46.80 & 37.4 & 49.80 & 14.41 & 6.73e--11 & 1.90e--09\\
*bn160220868 & 93 & 63.14 & 3.47 & 0.25 & 335.53 & 5.8 & 46.85 & 14.95 & 6.43e--11 & 1.74e--09\\
bn160215773 & 459 & 413.67 & 2.17 & -1.25 & 357.00 & 1.5 & 81.50 & 36.15 & 3.64e--11 & 1.19e--09\\
bn160211119 & 7 & 2.12 & 2.50 & 0.00 & 115.69 & 54.9 & 10.40 & 3.98 & 2.44e--09 & 7.57e--08\\
bn160206430 & 62 & 37.81 & 3.55 & 1.36 & 184.99 & 55.9 & 38.20 & 11.85 & 1.75e--10 & 4.22e--09\\
bn160118060 & 47 & 37.23 & 1.49 & -0.25 & 18.57 & 59.9 & 22.25 & 11.76 & 2.23e--10 & 6.91e--09\\
bn160113398 & 106 & 87.82 & 1.85 & 0.52 & 187.77 & 11.1 & 36.20 & 17.44 & 3.21e--11 & 9.09e--10\\
bn160107931 & 53 & 56.23 & -0.48 & ... & ... & ... & 10.00 & 14.22 & 6.95e--11 & 2.69e--09\\
bn160102500 & 83 & 55.78 & 3.36 & 1.15 & 218.82 & 7.5 & 43.25 & 14.17 & 1.08e--10 & 3.52e--09\\
bn151129333 & 100 & 72.76 & 2.99 & 0.46 & 66.53 & -12.0 & 44.75 & 15.97 & 1.28e--10 & 4.29e--09\\
bn151030999 & 862 & 877.21 & -0.53 & ... & ... & ... & 35.50 & 52.26 & 1.47e--12 & 2.70e--11\\
*bn151023104 & 11 & 2.91 & 3.50 & 0.35 & 13.04 & -10.2 & 14.89 & 4.38 & 1.53e--08 & 1.04e--06\\
*bn151022577 & 14 & 4.67 & 3.38 & -0.54 & 104.22 & 21.0 & 16.83 & 5.07 & 5.31e--10 & 1.73e--08\\
bn150928359 & 98 & 73.63 & 2.67 & -0.10 & 81.88 & 33.6 & 41.90 & 16.06 & 1.94e--10 & 8.55e--09\\
*bn150923297 & 7 & 1.48 & 3.14 & 0.62 & 323.04 & 27.4 & 11.05 & 3.61 & 4.38e--09 & 1.70e--07\\
bn150906944 & 18 & 9.79 & 2.26 & -1.11 & 215.54 & -1.1 & 16.37 & 6.68 & 8.31e--11 & 1.81e--09\\
*bn150904479 & 57 & 35.44 & 3.28 & 0.09 & 57.69 & -19.7 & 35.05 & 11.52 & 2.25e--10 & 7.53e--09\\
bn150811849 & 0 & 3.26 & ... & ... & ... & ... & 0.84 & 4.52 & 4.14e--11 & 9.99e--10\\
bn150721242 & 21 & 12.75 & 2.03 & 0.62 & 333.73 & 6.6 & 17.20 & 7.41 & 7.39e--10 & 2.86e--08\\
bn150705588 & 13 & 4.68 & 3.05 & -0.56 & 57.93 & -8.7 & 15.38 & 5.07 & 4.85e--10 & 1.58e--08\\
bn150622393 & 36 & 37.37 & -0.28 & ... & ... & ... & 9.70 & 11.79 & 1.27e--10 & 4.91e--09\\
bn150522944 & 15 & 5.63 & 3.17 & 0.14 & 120.60 & 51.3 & 16.89 & 5.39 & 3.81e--10 & 1.23e--08\\
bn150502435 & 131 & 133.06 & -0.21 & ... & ... & ... & 18.00 & 21.19 & 2.46e--11 & 8.24e--10\\
bn150416773 & 44 & 33.35 & 1.71 & -0.39 & 58.54 & 53.3 & 22.65 & 11.21 & 3.20e--10 & 9.91e--09\\
*bn150329288 & 64 & 38.48 & 3.71 & 0.89 & 166.54 & -9.7 & 40.00 & 11.93 & 2.07e--10 & 6.92e--09\\
*bn150318521 & 33 & 23.29 & 1.83 & -0.33 & 271.73 & -30.5 & 20.20 & 9.62 & 7.85e--10 & 1.03e--07\\
*bn150206407 & 16 & 7.05 & 2.80 & -0.02 & 208.73 & 56.5 & 16.95 & 5.89 & 8.22e--10 & 3.61e--08\\
*bn150201040 & 16 & 6.45 & 3.07 & -0.60 & 13.54 & 21.0 & 17.54 & 5.67 & 3.96e--10 & 1.28e--08\\
bn150131951 & 24 & 13.69 & 2.45 & -0.66 & 57.79 & 16.9 & 19.80 & 7.66 & 3.61e--10 & 1.21e--08\\
bn150128791 & 128 & 103.38 & 2.30 & -0.62 & 272.42 & 26.0 & 44.60 & 18.86 & 7.81e--11 & 2.62e--09\\
bn150126868 & 307 & 293.39 & 0.77 & ... & ... & ... & 43.50 & 30.75 & 2.03e--11 & 6.59e--10\\
bn150110433 & 70 & 68.29 & 0.17 & ... & ... & ... & 16.70 & 15.51 & 4.57e--10 & 1.11e--08\\
bn150105257 & 100 & 91.54 & 0.84 & ... & ... & ... & 26.00 & 17.81 & 8.76e--11 & 3.85e--09\\
bn141230142 & 106 & 74.23 & 3.43 & 1.58 & 57.48 & 4.3 & 49.80 & 16.12 & 2.73e--11 & 5.73e--10\\
bn141202470 & 7 & 2.42 & 2.26 & 0.03 & 145.89 & 60.7 & 10.11 & 4.16 & 1.26e--09 & 5.52e--08\\
\enddata
\tablecomments{Same as Table~\ref{tab:GBM_results_T90}, but using a search duration of 10 times $T_{90}$ for long GRBs and 20~s for short GRBs. There is not enough data for the analysis of bn160131174 and bn160515819 for this search duration.}
\tablenotetext{a}{($\rm{cm}^{-2}\rm{s}^{-1}\rm{keV}^{-1}$)}
\end{deluxetable*}

\section{Discussion and Outlook}
\begin{figure*}[htp]
 \centering
 \plottwo{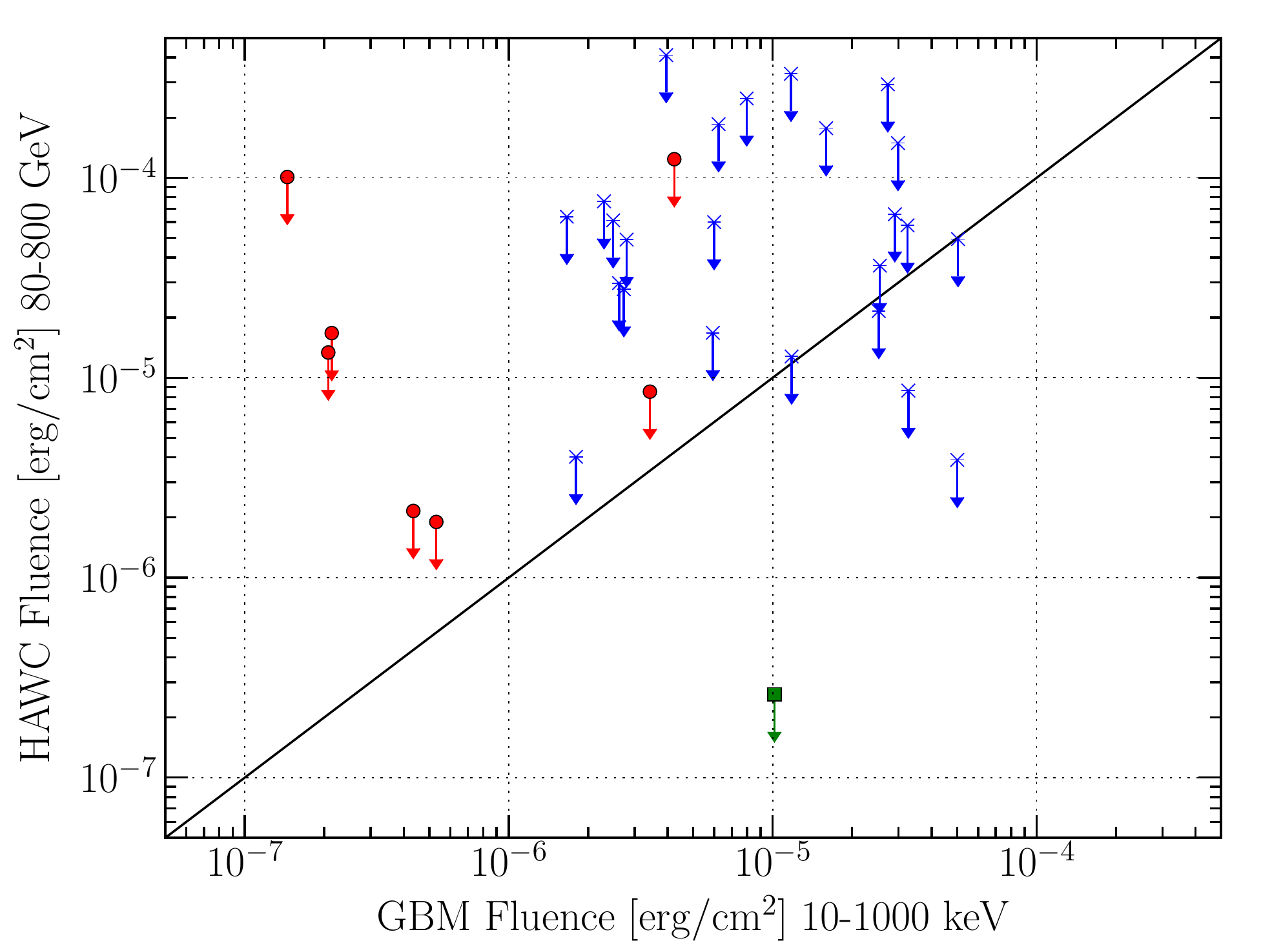}{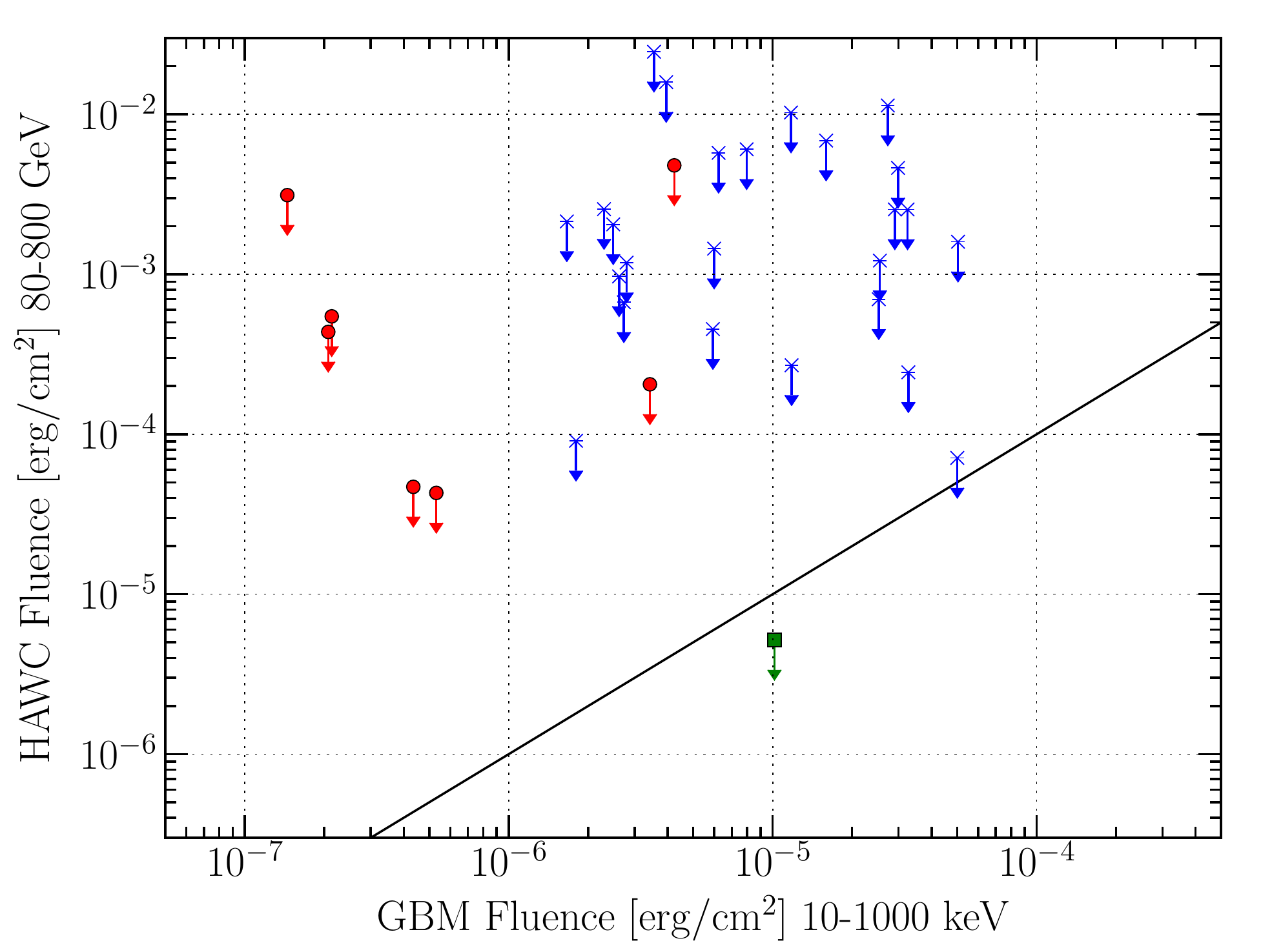}
 \caption{Comparison between the \emph{Fermi}-GBM fluence and the fluence implied by the HAWC upper limits obtained during the same time period for all GRBs completely inside the HAWC field of view for the two different redshifts (left panel $z=0.3$, right panel $z=1.0$). The circles (red) show short GRBs, the asterisks (blue) long GRBs, and the green square GRB~170206A. The black line shows an equal fluence in the \emph{Fermi}-GBM and HAWC energy range.}
 \label{fig:fluence_plot}
\end{figure*}

The two short GRBs 080825C and 090501 had ratios of high-energy (100~MeV--10~GeV) to low-energy (10~keV--1~MeV) fluences over 100\% \citep{bib:Fermi_LAT_GRB_catalogue}. Figure~\ref{fig:fluence_plot} compares the \emph{Fermi}-GBM fluence with the fluence implied by the HAWC upper limits for all GRB with an error radius that does not extend beyond the HAWC field of view. As expected, the fluence limits deteriorate when going to higher redshifts, since due to EBL absorption less signal is expected from a GRB. For all short GRBs that occurred during the first 18 months of HAWC data taking with the full HAWC detector, the HAWC fluence limits are above the fluence measured by \emph{Fermi}-GBM. For long GRBs, there are several GRBs (bn160215773, bn160113398, bn151030999, bn150126868) where the HAWC fluence limit is below the fluence measured by \emph{Fermi}-GBM if the GRB is as close as $z=0.3$. However, only about 20\% of long GRBs are closer than $z=1.0$, so the number of long GRBs at a redshift of 0.3 or closer should be fairly low.

\begin{figure}[htp]
 \centering
 \plotone{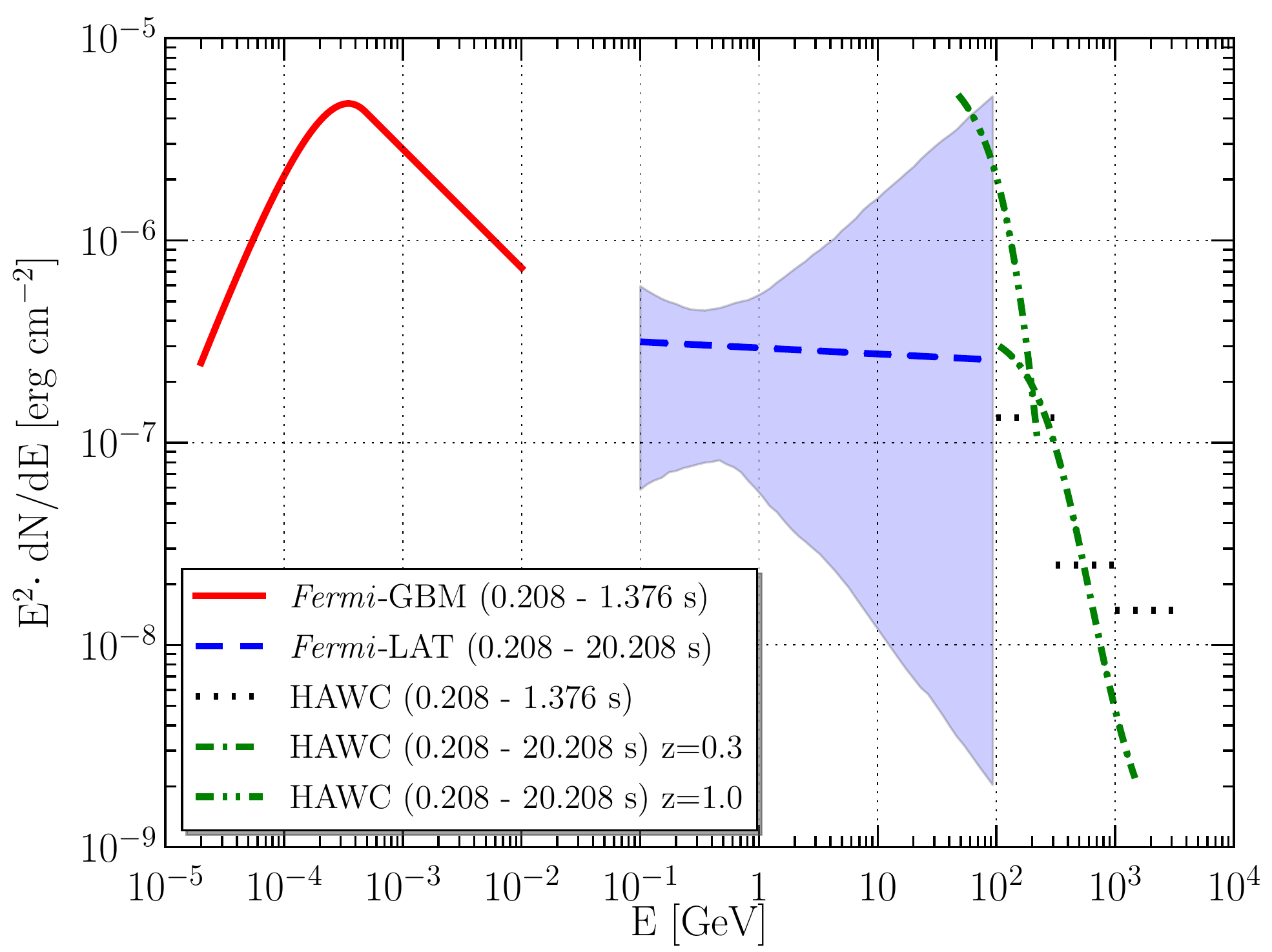}
 \caption{Solid line (red) shows the spectrum fitted to the prompt \emph{Fermi}-GBM data of GRB~170206, while dotted lines (black) show the ``quasi-differential'' limits assuming $E^{-2}$ obtained from the HAWC data taken during the same time period. Dashed line (blue) shows the best-fit spectrum obtained from the \emph{Fermi}-LAT data in the early afterglow and the shaded area the uncertainty taking into account the correlations and non-linearity of fit parameters. The green dashed–-dotted lines show the HAWC limits for two different assumed redshifts.}
 \label{fig:GRB170206_limits_plot}
\end{figure}

GRB~170206A is the only GRB where the fluence implied by the HAWC upper limits in the HAWC energy range is below the \emph{Fermi}-GBM fluence in the GBM energy range for both a redshift of 0.3 and 1, which is why this burst is the most interesting so far. Up to 2017 February, \emph{Swift} has detected 22 short GRBs that have a redshift measurement available, and only four of them were at a redshift larger than 1, so it is likely that GRB~170206A was at a redshift below 1. Taking the ratio implies that GRB~170206A had less than 3\% (50\%) of the fluence in the \emph{Fermi}-GBM energy range in the HAWC energy range for a burst at a redshift of 0.3 (1.0). If the GRB spectrum was cut off at 100~GeV, the ratio between HAWC and GBM would change to 63\% (130\%) for a burst at a redshift of 0.3 (1.0). A high-energy cut-off can be expected in SSC prompt emission models due to pair production. Figure~\ref{fig:GRB170206_limits_plot} compares the prompt spectrum measured by \emph{Fermi}-GBM to ``quasi-differential'' limits that are calculated by restricting an $E^{-2}$ spectrum to the energy ranges indicated by the limit bars.

Figure~\ref{fig:GRB170206_limits_plot} also shows the best-fit spectrum from \emph{Fermi}-LAT, obtained using the P8\_TRANSIENT010E class of LAT data. Events were selected with energies between 100 MeV and 100 GeV and a zenith angle smaller than $90\arcdeg$, contained within a circular region centered on the GRB position with a radius of $12\arcdeg$. The GRB was $67\arcdeg$ from the LAT boresight at the time of the trigger, at the edge of the LAT field of view, where the effective area is $\sim$30\% of the on-axis area. This suboptimal position resulted in only four detected photons in the 20~s following the trigger. The standard Fermi Science Tools v10r0p5 were used to perform an unbinned likelihood analysis with the tool \textit{gtlike}. A likelihood model containing a source with a power-law spectrum (the GRB), as well as the Galactic and the Isotropic templates provided by the \emph{Fermi}-LAT collaboration, were adopted. The normalization of the Galactic template was kept fixed. As customary for LAT analysis, the likelihood-ratio test \citep{bib:likelihood_ratio_test} was used to assess whether the GRB is detected or not. It uses as test statistic ${\rm TS} = 2(\log{L_{1}} - \log{L_{0}})$, where $L_{1}$ and $L_{0}$ are the maximum of the likelihood for a model with and without the GRB, respectively. Monte Carlo simulations show that TS is distributed as $1/2~\chi^2$ with 1 degree of freedom \citep{bib:EGRET_TS}. ${\rm TS} = 28$ was found for the GRB, corresponding to a significance of $5.3\sigma$. The best-fit photon index is $-2.0 \pm 0.6$, with an average flux (100~MeV--100~GeV) of $(9\pm6)\times 10^{-4}$~ph~cm$^{-2}$~s$^{-1}$. HAWC limits can exclude a very hard photon index or require a spectral break in between the highest energy LAT photon (811~MeV) and the HAWC energy range for a nearby burst ($z=0.3$). For a GRB at a redshift of $z=1.0$, the HAWC limits are only marginally constraining.

GRB~160310A \citep{bib:GRB160310A} is the other well-located LAT GRB analysed in this paper, but the zenith angle of $34\arcdeg$ is far from optimal for HAWC. Furthermore, this GRB is not detected in LAT data using the same selection and technique as used for GRB~170206A in the time intervals considered in this paper. It is detected on a much longer time scale (0--10~ks), as also reported in \citet{bib:GRB160310A}. Thus, in this case it is not possible to directly compare the HAWC upper limits to the LAT spectrum as was done for GRB 170206A.

The previous estimate of the GRB detection rate in HAWC used only triggers by \emph{Fermi}-GBM, while in the current paper those GRBs as well as GRBs triggering \emph{Swift} were analysed. \emph{Swift} added about 15\% uniquely identified bursts, so the GRB detection rate of HAWC should also be slightly higher. However, in \cite{bib:HAWC_GRB_rate}, a trigger threshold of 30 PMTs was used, which is significantly lower than the threshold of the current analysis (6.7\% of PMTs participating in the event). Thus, the non-detection of a GRB within one and a half years of operations is not in conflict with previous estimates.

This paper has presented upper limits for VHE emission from GRBs observed during the first one and a half years of the HAWC Gamma-ray Observatory. None of the bursts was significantly detected. If a SSC component is present in GRB~170206A, the HAWC upper limits constrain the expected cut-off to be less than $100~\rm{GeV}$ for reasonable assumptions about the energetics and redshift of the burst. With the current statistics, the limits on the GRB detection rate in HAWC still do not allow strong conclusions about the distribution of the high-energy photon index, high-energy cut-offs or the fraction of GRBs with additional power-law components to be drawn. In order for the limits to become constraining to physical models, more years of operation or a more sensitive analysis are needed. For bursts at high redshift, it is essential to expand the analysis to trigger threshold, ideally with a set of reliable GH separation cuts. It will lower the energy reach of HAWC and thereby improve HAWC's sensitivity to GRBs. The current search is most sensitive to emission that starts close to the trigger time on the time scales searched. Searches on other time scales and/or start times or a more model-dependent search, e.g. assuming a certain light-curve shape, are possible future avenues to explore. Another possible improvement to the analysis of GBM burst is to search the error ellipse with a higher containment probability and to take systematic uncertainties into account.

\begin{acknowledgements}
We acknowledge the support from: the US National Science Foundation (NSF); the US Department of Energy Office of High-Energy Physics; the Laboratory Directed Research and Development (LDRD) program of Los Alamos National Laboratory; NASA Fermi GI (grant 81206); Consejo Nacional de Ciencia y Tecnolog\'{\i}a (CONACyT), M{\'e}xico (grants 271051, 232656, 260378, 179588, 239762, 254964, 271737, 258865, 243290, 132197), Laboratorio Nacional HAWC de rayos gamma; L'OREAL Fellowship for Women in Science 2014; Red HAWC, M{\'e}xico; DGAPA-UNAM (grants IG100317, IN111315, IN111716-3, IA102715, 109916, IA102917); VIEP-BUAP; PIFI 2012, 2013, PROFOCIE 2014, 2015; the University of Wisconsin Alumni Research Foundation; the Institute of Geophysics, Planetary Physics, and Signatures at Los Alamos National Laboratory; Polish Science Centre grant DEC-2014/13/B/ST9/945; Coordinaci{\'o}n de la Investigaci{\'o}n Cient\'{\i}fica de la Universidad Michoacana. Thanks to Luciano D\'{\i}az and Eduardo Murrieta for technical support.

G. Vianello is a member of the LAT Collaboration. The \textit{Fermi}-LAT Collaboration acknowledges generous ongoing support from a number of agencies and institutes that have supported both the development and the operation of the LAT as well as scientific data analysis. These include the National Aeronautics and Space Administration and the Department of Energy in the United States; the Commissariat \`a l'Energie Atomique and the Centre National de la Recherche Scientifique / Institut National de Physique Nucl\'eaire et de Physique des Particules in France; the Agenzia Spaziale Italiana and the Istituto Nazionale di Fisica Nucleare in Italy; the Ministry of Education, Culture, Sports, Science and Technology (MEXT), High Energy Accelerator Research Organization (KEK) and Japan Aerospace Exploration Agency (JAXA) in Japan; and the K.~A.~Wallenberg Foundation, the Swedish Research Council and the Swedish National Space Board in Sweden. Additional support for science analysis during the operations phase is gratefully acknowledged from the Istituto Nazionale di Astrofisica in Italy and the Centre National d'\'Etudes Spatiales in France.
\end{acknowledgements}

\software{ROOT \citep{bib:ROOT}, GammaPy \citep{bib:gammapy}, Matplotlib \citep{bib:matplotlib}, NumPy \citep{bib:numpy}, SciPy \citep{bib:scipy}}

\bibliography{references}

\appendix
\section{GRB Selection}
GRBs occurring in the field of view of HAWC (down to a zenith angle of 51\arcdeg) are found using the online tables for \emph{Fermi}-LAT\footnote{\url{http://fermi.gsfc.nasa.gov/ssc/observations/types/grbs/lat_grbs/}}, \emph{Fermi}-GBM\footnote{\url{https://heasarc.gsfc.nasa.gov/W3Browse/all/fermigbrst.html}} and \emph{Swift}\footnote{\url{http://swift.gsfc.nasa.gov/archive/grb\_table/}} (Tables~\ref{tab:LAT_GRBs}--\ref{tab:GBM_GRBs}). The GRBOX: Gamma-Ray Burst Online Index\footnote{\url{http://www.astro.caltech.edu/grbox/grbox.php}} was additionally searched for GRBs from other experiments, but no matches were found.

For the \emph{Fermi}-LAT-detected GRBs, the trigger time, $T_{90}$ and $T_{90}$ Start, are taken from the \emph{Fermi}-GBM catalogue. For GRB~170206A, the position and error stem from the best on-ground location of the LAT detector \citep{bib:170206A_LAT}. For GRB~160310A, the position is taken from NOT optical observations \citep{bib:160310A_position}. GRB~150416A was only located by \emph{Fermi}-GBM, so its position and error are taken from the \emph{Fermi}-GBM catalogue.

The \emph{Swift} position is, in order of preference, the UVOT, XRT, or BAT position. For the \emph{Swift} / \emph{Fermi}-GBM co-detection, the burst is analysed using the \emph{Swift} information.

\begin{deluxetable}{l|ccccccc}
\tabletypesize{\footnotesize}
\tablecolumns{7}
\tablewidth{0pt}
\tablecaption{Data of LAT-detected GRBs \label{tab:LAT_GRBs}}
\tablehead{\vspace{-0.1cm} & \colhead{Time} & \colhead{R.A. J2000} & \colhead{Decl. J2000} & \colhead{Error} & \colhead{Zen} & \colhead{$T_{90}$} &
           \colhead{$T_{90}$ Start} \\
           \vspace{-0.1cm} & \colhead{UTC}  & \colhead{}         & \colhead{}          & \colhead{(deg)}   & \colhead{(deg)} & \colhead{(s)} & \colhead{(s)}}
\startdata
170206A & 10:51:57.696 & 14h11m09.6s  & +14d28m48.0s & 0.85     & 11.1 & 1.168  & 0.208 \\
160310A & 00:22:58.468 & 06h35m17.33s & -07d12m56.0s & 1.389E-4 & 34.3 & 18.173 & 0.003 \\
150416A & 18:33:25.965 & 03h55m00.0s  & +52d57m36s   & 1.9300   & 42.7 & 33.281 & 0.512 \\
\enddata
\tablecomments{The three GRBs detected by LAT that are analysed in this paper.}
\end{deluxetable}

\begin{deluxetable}{l|ccccccc}
\tabletypesize{\footnotesize}
\tablecolumns{8}
\tablewidth{0pt}
\tablecaption{Data of \emph{Swift}-detected GRBs \label{tab:Swift_GRBs}}
\tablehead{\vspace{-0.1cm} & \colhead{Trigger} & \colhead{Time} & \colhead{R.A. J2000} & \colhead{Decl. J2000} & \colhead{Zenith} & \colhead{BAT $T_{90}$} & \colhead{Redshift}\\
           \vspace{-0.1cm} & \colhead{Number}  & \colhead{UTC}  & \colhead{}         & \colhead{}          & \colhead{(deg)}    & \colhead{(s)}
          & \colhead{}}
\startdata
160410A & 682269 & 05:09:48 & 10h02m44.37s & 03d28m42.7s & 31.6 & 8.2 & 1.717\tablenotemark{a} \\
151228B & 668641 & 22:47:14 & 22h57m41.87s & 08d04m53.7s & 11.2 & 48.0 & n/a \\
151205A & 666352 & 15:46:00 & 15h17m09.27s & 35d44m38.4s & 21.9 & 62.8 & n/a \\
150817A & 652334 & 02:05:13 & 16h38m31.47s & -12d03m10.6s & 32.4 & 38.8 & n/a \\
150811A & 651882 & 04:06:09 & 19h25m21.40s & -15d25m31.1s & 35.3 & 34.00 & n/a \\
150716A & 649157 & 07:06:43 & 18h33m57.10s & -12d58m48.9s & 40.0 & 44 & n/a \\
150710A & 648437 & 00:28:02 & 12h57m52.91s & 14d19m05.0s & 5.4 & 0.15 & n/a \\
150530B & GA     & 13:28:38 & 00h29m59.04s & 44d17m24.0s & 28.3 & n/a & n/a \\
150530A & 642018 & 11:42:18 & 21h50m02.94s & 57d30m59.8s & 38.6 & 6.62 & n/a \\
150527A & 641698 & 06:48:55 & 19h15m50.32s & 04d12m06.9s & 41.5 & 112 & n/a \\
150428A & 639275 & 01:30:40 & 12h34m09.22s & 06d57m13.2s & 47.9 & 53.2 & n/a \\
150423A & 638808 & 06:28:04 & 14h46m18.96s & 12d17m00.6s & 12.6 & 0.22 & 1.394\tablenotemark{b} \\
150323B & GA     & 09:28:38 & 17h21m48.83s & 38d19m04.6s & 36.1 & $\sim$60 & n/a \\
150317A & 635148 & 04:22:42 & 09h15m56.32s & 55d27m56.6s & 36.5 & 23.29 & n/a \\
150302A & 633237 & 05:42:36 & 11h42m07.48s & 36d48m39.8s & 30.0 & 23.74 & n/a \\
150211A & 630714 & 11:52:11 & 16h59m26.21s & 55d23m36.2s & 44.0 & 13.6 & n/a \\
150203A & 629578 & 04:09:07 & 06h33m35.80s & 06d57m13.4s & 12.1 & 25.8 & n/a \\
141205A & GA     & 08:05:17 & 06h11m26.16s & 37d52m33.6s & 19.4 & 1.1 & n/a \\
\enddata
\tablecomments{The 18 \emph{Swift}-detected GRBs in the HAWC field of view that are analysed in this paper. GA means the burst was found in a ground analysis. For GRB~150530B, a $T_{90}$ of 20~s is assumed.}
\tablenotetext{a}{\cite{bib:160410A_redshift}}
\tablenotetext{b}{\cite{bib:150423A_redshift}, redshift only tentative}
\end{deluxetable}

\begin{deluxetable}{l|ccccccc}
\tabletypesize{\footnotesize}
\tablecolumns{8}
\tablewidth{0pt}
\tablecaption{Data of \emph{Fermi}-GBM-detected GRBs \label{tab:GBM_GRBs}}
\tablehead{\vspace{-0.1cm} & \colhead{Time} & \colhead{R.A. J2000} & \colhead{Decl. J2000} & \colhead{Error} & \colhead{Zen} & \colhead{GBM $T_{90}$} &
           \colhead{$T_{90}$ Start} \\
           \vspace{-0.1cm} & \colhead{UTC}  & \colhead{}         & \colhead{}          & \colhead{(deg)}   & \colhead{(deg)} & \colhead{(s)} & \colhead{(s)}}
\startdata
bn160605847  & 20:19:29.764 & 07h23m55.2s & -19d49m48s & 4.9500 & 39.8 & 5.504 & -0.064 \\
bn160527080  & 01:55:37.323 & 14h33m43.2s & +06d40m48s & 6.5500 & 42.6 & 25.344 & -18.688 \\
bn160521839  & 20:07:53.873 & 04h50m02.4s & -13d42m00s & 4.4400 & 34.7 & 15.872 & -0.768 \\
bn160515819  & 19:38:41.037 & 05h22m16.8s & -16d18m36s & 3.4300 & 36.5 & 84.481 & -0.768 \\
bn160406503  & 12:04:36.798 & 17h27m11.3s & +32d15m47s & 11.8000 & 20.3 & 0.432 & -0.336 \\
bn160315739  & 17:44:50.341 & 21h19m38.4s & -22d06m36s & 23.0500 & 46.6 & 3.328 & -2.816 \\
bn160303201  & 04:49:32.084 & 10h53m57.6s & +56d56m24s & 9.0500 & 42.9 & 48.129 & 1.792 \\
bn160301215  & 05:10:18.519 & 07h37m36.0s & +02d17m24s & 3.4800 & 29.7 & 29.697 & 0.256 \\
bn160228034  & 00:48:52.491 & 02h08m50.4s & +39d22m48s & 12.4300 & 39.8 & 16.128 & -12.544 \\
bn160220868  & 20:50:12.110 & 21h47m33.6s & +06d03m00s & 15.7800 & 39.5 & 22.528 & -7.424 \\
bn160215773  & 18:33:30.387 & 23h47m09.6s & +01d43m48s & 3.4400 & 34.6 & 141.314 & 54.273 \\
bn160211119  & 02:50:48.276 & 08h12m48.0s & +53d25m48s & 4.9700 & 44.9 & 0.960 & -0.768 \\
bn160206430  & 10:19:12.431 & 12h17m02.4s & +52d24m36s & 4.1700 & 34.1 & 21.504 & -5.632 \\
bn160131174  & 04:09:56.714 & 07h32m07.2s & +15d29m24s & 4.7100 & 17.8 & 205.315 & -0.768 \\
bn160118060  & 01:25:42.450 & 01h10m52.8s & +59d46m48s & 1.4900 & 44.1 & 46.849 & 1.024 \\
bn160113398  & 09:32:30.524 & 12h29m02.4s & +11d31m48s & 1.2000 & 29.2 & 24.576 & 26.176 \\
bn160107931  & 22:20:41.502 & 19h58m40.8s & +06d24m47s & 0.1700 & 45.4 & 113.922 & -11.520 \\
bn160102500  & 11:59:22.628 & 14h55m04.8s & +06d22m48s & 5.8100 & 40.8 & 25.344 & -12.032 \\
bn151129333  & 08:00:06.085 & 04h03m16.8s & -11d29m24s & 5.6700 & 42.2 & 52.224 & -32.768 \\
bn151030999  & 23:58:22.637 & 19h50m31.2s & +30d51m00s & 1.0000 & 12.3 & 116.482 & 6.144 \\
bn151023104  & 02:29:25.137 & 23h58m57.6s & -17d09m00s & 16.4000 & 45.8 & 10.240 & -2.304 \\
bn151022577  & 13:51:02.089 & 07h21m28.8s & +40d13m48s & 21.3600 & 33.6 & 0.320 & -0.160 \\
bn150928359  & 08:37:19.023 & 05h35m07.2s & +34d14m24s & 4.5700 & 42.8 & 53.504 & -30.208 \\
bn150923297  & 07:07:36.184 & 21h07m12.0s & +31d49m12s & 10.7600 & 50.3 & 0.192 & -0.112 \\
bn150906944  & 22:38:47.307 & 14h08m09.6s & +01d05m24s & 5.1900 & 23.8 & 0.320 & -0.256 \\
bn150904479  & 11:30:20.956 & 04h28m19.2s & -20d22m48s & 10.8900 & 40.2 & 23.296 & -4.096 \\
bn150811849  & 20:22:13.749 & 12h25m23.5s & -14d06m18s & 0.9900 & 37.8 & 0.640 & -0.064 \\
bn150721242  & 05:49:08.934 & 22h16m33.6s & +07d45m36s & 1.5000 & 45.7 & 18.432 & 1.024 \\
bn150705588  & 14:07:11.608 & 04h26m09.6s & -06d37m12s & 12.6000 & 38.3 & 0.704 & -0.256 \\
bn150622393  & 09:26:32.023 & 17h48m12.0s & +33d15m00s & 1.0000 & 44.4 & 60.673 & 1.024 \\
bn150522944  & 22:38:44.068 & 08h43m26.4s & +58d34m48s & 10.4700 & 40.0 & 1.024 & -0.128 \\
bn150502435  & 10:25:55.163 & 16h05m50.4s & +42d03m36s & 1.0000 & 39.2 & 109.314 & 5.888 \\
bn150329288  & 06:55:19.123 & 10h52m00.0s & -12d19m12s & 11.7300 & 42.9 & 28.928 & -2.560 \\
bn150318521  & 12:29:53.024 & 17h56m12.0s & -30d13m12s & 2.1900 & 49.3 & 94.720 & 0.768 \\
bn150206407  & 09:46:27.485 & 14h42m21.6s & +57d30m00s & 6.7600 & 46.3 & 5.120 & -1.024 \\
bn150201040  & 00:56:54.289 & 00h22m31.2s & +19d45m00s & 13.9200 & 39.5 & 0.512 & -0.320 \\
bn150131951  & 22:49:26.183 & 04h09m04.8s & +19d16m12s & 5.3100 & 43.9 & 8.192 & -0.256 \\
bn150128791  & 18:59:14.294 & 18h09m04.8s & +27d46m48s & 3.3200 & 40.0 & 85.248 & -3.328 \\
bn150126868  & 20:50:35.778 & 23h22m00.7s & -12d22m05s & 0.5100 & 32.7 & 96.513 & 6.400 \\
bn150110433  & 10:23:38.232 & 14h28m00.0s & +18d54m00s & 1.0000 & 46.1 & 74.304 & -1.344 \\
bn150105257  & 06:10:00.463 & 08h17m16.8s & -14d46m48s & 1.0000 & 41.7 & 73.729 & 3.072 \\
bn141230142  & 03:24:22.637 & 03h47m55.2s & +01d35m24s & 3.8600 & 18.0 & 9.856 & 0.000 \\
bn141202470  & 11:17:05.606 & 09h40m02.4s & +59d52m12s & 3.3200 & 40.8 & 1.344 & -0.064 \\
\enddata
\tablecomments{The 43 GRBs detected only by \emph{Fermi}-GBM in the HAWC field of view that are analysed in this paper.}
\end{deluxetable}

\end{document}